\documentclass[12pt,a4paper]{iopart}
\usepackage[T1]{fontenc}
\usepackage[utf8]{inputenc}

\usepackage[usenames, dvipsnames]{xcolor}
\definecolor{linkcolor}{rgb}{0.0,0.3,0.5}
\usepackage[unicode,colorlinks=true,linkcolor=linkcolor,urlcolor=linkcolor,citecolor=linkcolor,pdfusetitle]{hyperref}
\usepackage{cite}
\usepackage[all]{hypcap}
\usepackage{graphicx}
\usepackage{xspace}
\usepackage{mathrsfs}
\usepackage{pifont} %
\usepackage{lmodern}
\usepackage{printlen}
\usepackage{pdflscape}
\usepackage{layout}
\usepackage{verbatim}
\usepackage{listings}
\usepackage{multicol}

\usepackage{iopams}
\usepackage{orcidlink}

\expandafter\let\csname equation*\endcsname\relax
\expandafter\let\csname endequation*\endcsname\relax

\usepackage{amsmath,amsfonts,amssymb}
\usepackage{amsthm}
\usepackage{etoolbox}
\usepackage{relsize, exscale}

\usepackage{longtable}

\usepackage{microtype}
\usepackage[normalem]{ulem} %

\newcommand{\numberofsimulations}{3,756\xspace}
\newcommand{\mediannumberoforbits}{22\xspace}
\newcommand{\largestnumberoforbits}{148\xspace}
\newcommand{\totalnumberoflevs}{10,557\xspace}
\newcommand{\averagenumberoflevs}{2.8\xspace}
\newcommand{\oldtotalsize}{12~TiB\xspace}
\newcommand{\oldwaveformsize}{180~GiB\xspace}
\newcommand{\oldnumberofsimulations}{2,018\xspace}
\newcommand{\newtotalsize}{410~GiB\xspace}
\newcommand{\newwaveformsize}{14~GiB\xspace}

\newcommand{\bestwaveformratio}{23\xspace}

\graphicspath{{../Scripts/}{../Scripts/catalog_overview/}{../Scripts/waveform_comparison/}{../Scripts/compression_ratio_v_tolerance/}{../Scripts/compression_error/}{../Scripts/showoff/}}

\newcommand{\simscharged}{480,000,000\xspace}

\newcommand{\macrocolor}[1]{\textcolor{red}{#1}}
\renewcommand{\macrocolor}[1]{#1}

\newcommand{\XOR}{\texttt{XOR}\xspace}

\newcommand{\software}[1]{\texttt{#1}}

\input{dimensions}

\begin{document}

\title{The SXS Collaboration's third catalog\\ of binary black hole simulations}
\newcommand{\AEI}{
  Max Planck Institute for Gravitational Physics
  (Albert Einstein Institute),
  Am M{\"u}hlenberg 1,
  14476 Potsdam, Germany}%
\newcommand{\BHam}{School of Physics and Astronomy and Institute for
  Gravitational Wave Astronomy, University of Birmingham, Edgbaston,
  Birmingham, B15 9TT, UK}
\newcommand{\Caltech}{Theoretical Astrophysics 350-17,
    California Institute of Technology, Pasadena, CA 91125, USA}
\newcommand{\CENTRA}{CENTRA, Departamento de F\'{\i}sica, Instituto Superior T\'ecnico, Universidade de Lisboa, Avenida Rovisco Pais 1, 1049-001 Lisboa, Portugal}
\newcommand{\Cornell}{Cornell Center for Astrophysics and Planetary Science,
    Cornell University, Ithaca, New York 14853, USA}
\newcommand{\CITA}{Canadian Institute for Theoretical
    Astrophysics, 60 St.~George Street, University of Toronto,
    Toronto, ON M5S 3H8, Canada} %
\newcommand{\JPL}{Jet Propulsion Laboratory, California Institute of Technology, Pasadena, CA
  91109, USA}
\newcommand{\KITP}{Kavli Institute for Theoretical Physics, University of California Santa Barbara, Kohn Hall, Lagoon Rd, Santa Barbara, CA 93106, USA}
\newcommand{\Manchester}{University of Manchester, Manchester, UK}
\newcommand{\MIT}{Department of Physics and MIT Kavli Institute, Cambridge, MA 02139, USA}
\newcommand{\Pullman}{Department of Physics and Astronomy, Washington State
  University, Pullman, Washington 99164, USA}
\newcommand{\UMiss}{Department of Physics and Astronomy,
    University of Mississippi, University, MS 38677, USA}
\newcommand{\Radboud}{Department of Astrophysics/IMAPP, Radboud University Nijmegen, P.O. Box
9010, 6500 GL Nijmegen, The Netherlands}
\newcommand{\TorontoPhysics}{Department of Physics
  60 St.~George Street, University of Toronto,
    Toronto, ON M5S 3H8, Canada} %
\newcommand{\Tokyo}{Research Center for the Early Universe, University of Tokyo, Tokyo, 113-0033, Japan}%
\newcommand{\GWPAC}{Nicholas and Lee Begovich Center for Gravitational-Wave Physics and
    Astronomy, California State University Fullerton,
    Fullerton, California 92834, USA} %
\newcommand{\ChristopherNewport}{Christopher Newport University, Newport News, VA 23606, USA}
\newcommand{\UMass}{Department of Mathematics,
Center for Scientific Computing and Data Science Research,
University of Massachusetts, Dartmouth, MA 02747, USA
} %
\newcommand{\UTA}{Theory Group, Department of Physics, University of Texas at Austin, Austin, TX 78712, USA}
\newcommand{\Icts}{International Centre for Theoretical Sciences, Tata Institute
  of Fundamental Research, Bangalore 560089, India}
\newcommand{\Coimbra}{CFisUC, Department of Physics, University of
    Coimbra, 3004-516 Coimbra, Portugal}
\newcommand{\Balears}{Departament de Física, Universitat de les Illes Balears,
  IAC3 -- IEEC, Crta.~Valldemossa km 7.5, E-07122 Palma, Spain}
\newcommand{\KingsLondon}{Theoretical Particle Physics and Cosmology Group,
  Physics Department, King's College London, Strand, London WC2R 2LS, United
  Kingdom}
\newcommand{\VSM}{{Department of Physics, Vivekananda Satavarshiki
  Mahavidyalaya (affiliated to Vidyasagar University), Manikpara 721513, West
  Bengal, India}}
\newcommand{\Oberlin}{Department of Physics and Astronomy, Oberlin College}
\newcommand{\Wigner}{HUN-REN Wigner RCP, H-1121 Budapest, Konkoly Thege Mikl\'{o}s \'{u}t  29-33, Hungary}

\newcommand{\todo}[1]{{\color{blue}\textsf{[TODO:#1]}}}

\newcommand{\NsimsMoreThanOneRes}{\macrocolor{1872}}
\newcommand{\NsimsMoreThanTwoRes}{\macrocolor{1777}}
\newcommand{\NsimsTotal}{\macrocolor{1872}}
\newcommand{\NmismatchesTotal}{\macrocolor{37440}}
\newcommand{\NmismatchesNonconvergentH}{\macrocolor{3558}}
\newcommand{\NmismatchesNonconvergentPsi}{\macrocolor{4851}}

\def\aj{Astron.~J.}
\def\apj{Astrophys.~J.}
\def\apjl{Astrophys.~J.~Lett.}
\def\apjs{Astrophys.~J.~Supp.~Ser.}
\def\aa{Astron.~Astrophys.}
\def\aap{Astron.~Astrophys.}
\def\aapr{Astro.~Astrophys.~Rev.}
\def\araa{Ann.~Rev.~Astron.~Astroph.}
\def\apss{Astrophys.~Space~Sci.}
\def\aaps{Astron.~Astrophys.~Suppl.~Ser.}
\def\cqg{Class.~Quantum~Grav.}
\def\physrep{Phys.~Rep.}
\def\mnras{Mon.~Not.~Roy.~Astron.~Soc.}
\def\nat{Nature}
\def\prl{Phys.~Rev.~Lett.}
\def\prd{Phys.~Rev.~D.}
\def\prc{Phys.~Rev.~C.}
\def\azh{Soviet Astron.}
\def\jqsrt{JQSRT}

\newlength{\savedmathindent}
\setlength{\savedmathindent}{\mathindent}
\setlength{\mathindent}{0pc}

\author{%
Mark~A.~Scheel~\orcidlink{0000-0001-6656-9134}$^{1}$,
Michael~Boyle~\orcidlink{0000-0002-5075-5116}$^{2}$,
Keefe~Mitman~\orcidlink{0000-0003-0276-3856}$^{2}$,
Nils~Deppe~\orcidlink{0000-0003-4557-4115}$^{2}$,
Leo~C.~Stein~\orcidlink{0000-0001-7559-9597}$^{3}$,
Crist\'obal~Armaza~\orcidlink{0000-0002-1791-0743}$^{2}$,
Marceline~S.~Bonilla~\orcidlink{0000-0003-4502-528X}$^{4}$,
Luisa~T.~Buchman$^{5}$,
Andrea~Ceja~\orcidlink{0000-0002-1681-7299}$^{4}$,
Himanshu~Chaudhary~\orcidlink{0000-0002-4101-0534}$^{1}$,
Yitian~Chen~\orcidlink{0000-0002-8664-9702}$^{2}$,
Maxence~Corman~\orcidlink{0000-0003-2855-1149}$^{6}$,
Károly~Zoltán~Csukás~\orcidlink{0000-0002-2408-1103}$^{7}$,
C.~Melize~Ferrus~\orcidlink{0000-0002-2842-2067}$^{8}$,
Scott~E.~Field~\orcidlink{0000-0002-6037-3277}$^{9}$,
Matthew~Giesler~\orcidlink{0000-0003-2300-893X}$^{2}$,
Sarah~Habib~\orcidlink{0000-0002-4725-4978}$^{1}$,
Fran\c{c}ois~H\'{e}bert~\orcidlink{0000-0001-9009-6955}$^{1}$,
Daniel~A.~Hemberger$^{1}$,
Dante~A.~B.~Iozzo~\orcidlink{0000-0002-7244-1900}$^{2}$,
Tousif~Islam~\orcidlink{0000-0002-3434-0084}$^{10,9}$,
Ken~Z.~Jones~\orcidlink{0009-0003-1034-0498}$^{4}$,
Aniket~Khairnar~\orcidlink{0000-0001-5138-572X}$^{3}$,
Lawrence~E.~Kidder~\orcidlink{0000-0001-5392-7342}$^{2}$,
Taylor~Knapp~\orcidlink{0000-0001-8474-4143}$^{1}$,
Prayush~Kumar~\orcidlink{0000-0001-5523-4603}$^{11}$,
Guillermo~Lara~\orcidlink{0000-0001-9461-6292}$^{6}$,
Oliver~Long~\orcidlink{0000-0002-3897-9272}$^{6}$,
Geoffrey~Lovelace~\orcidlink{0000-0002-7084-1070}$^{4,1}$,
Sizheng~Ma~\orcidlink{0000-0002-4645-453X}$^{1}$,
Denyz~Melchor~\orcidlink{0000-0002-7854-1953}$^{4}$,
Marlo~Morales~\orcidlink{0000-0002-0593-4318}$^{4}$,
Jordan~Moxon~\orcidlink{0000-0001-9891-8677}$^{1}$,
Peter~James~Nee~\orcidlink{0000-0002-2362-5420}$^{6}$,
Kyle~C.~Nelli\orcidlink{0000-0003-2426-8768}$^{1}$,
Eamonn~O'Shea$^{2}$,
Serguei~Ossokine~\orcidlink{0000-0002-2579-1246}$^{6}$,
Robert~Owen~\orcidlink{0000-0002-1511-4532}$^{12}$,
Harald~P.~Pfeiffer~\orcidlink{0000-0001-9288-519X}$^{6}$,
Isabella~G.~Pretto~\orcidlink{0009-0001-7552-551X}$^{1}$,
Teresita~Ramirez-Aguilar~\orcidlink{0000-0003-0994-115X}$^{4}$,
Antoni~Ramos-Buades~\orcidlink{0000-0002-6874-7421}$^{13}$,
Adhrit~Ravichandran~\orcidlink{0000-0002-9589-3168}$^{9}$,
Abhishek~Ravishankar~\orcidlink{0009-0006-6519-8996}$^{9}$,
Samuel~Rodriguez~\orcidlink{0000-0002-1879-8810}$^{4}$,
Hannes~R.~R\"uter~\orcidlink{0000-0002-3442-5360}$^{14}$,
Jennifer~Sanchez~\orcidlink{0000-0002-5335-4924}$^{4}$,
Md~Arif~Shaikh~\orcidlink{0000-0003-0826-6164}$^{15}$,
Dongze~Sun~\orcidlink{0000-0003-0167-4392}$^{1}$,
B\'ela~Szil\'agyi$^{1}$,
Daniel~Tellez~\orcidlink{0009-0008-7784-2528}$^{4}$,
Saul~A.~Teukolsky~\orcidlink{0000-0001-9765-4526}$^{2,1}$,
Sierra~Thomas~\orcidlink{0000-0003-3574-2090}$^{4}$,
William~Throwe~\orcidlink{0000-0001-5059-4378}$^{2}$,
Vijay~Varma~\orcidlink{0000-0002-9994-1761}$^{9}$,
Nils~L.~Vu~\orcidlink{0000-0002-5767-3949}$^{1}$,
Marissa~Walker~\orcidlink{0000-0002-7176-6914}$^{4,16}$,
Nikolas~A.~Wittek~\orcidlink{0000-0001-8575-5450}$^{6}$
and
Jooheon~Yoo~\orcidlink{0000-0002-3251-0924}$^{2}$
}

\newcommand{\affilInfo}{%
\address{$^{1}$~\Caltech}
\address{$^{2}$~\Cornell}
\address{$^{3}$~\UMiss}
\address{$^{4}$~\GWPAC}
\address{$^{5}$~\Pullman}
\address{$^{6}$~\AEI}
\address{$^{7}$~\Wigner}
\address{$^{8}$~\KingsLondon}
\address{$^{9}$~\UMass}
\address{$^{10}$~\KITP}
\address{$^{11}$~\Icts}
\address{$^{12}$~\Oberlin}
\address{$^{13}$~\Balears}
\address{$^{14}$~\CENTRA}
\address{$^{15}$~\VSM}
\address{$^{16}$~\ChristopherNewport}
}

\begin{indented}\item[]{} (Affiliation list at end.)\end{indented}

\hypersetup{pdfauthor={Scheel et al.}}

\begin{abstract}
We present a major update to the Simulating eXtreme Spacetimes (SXS)
Collaboration's catalog of binary black hole simulations. Using highly
efficient spectral methods implemented in the Spectral Einstein Code (\software{SpEC}), we
have nearly doubled the total number of binary configurations from \oldnumberofsimulations
to \numberofsimulations. The catalog now densely covers the parameter space
with precessing simulations up to mass ratio
$q=8$ and dimensionless spins up to
$|\vec{\chi}|\le0.8$ with near-zero eccentricity. The catalog also includes
some simulations at higher mass
ratios with moderate spin and more than 250 eccentric
simulations. We have also deprecated and rerun some
simulations from our previous catalog (e.g., simulations run with a much older 
version of \software{SpEC} or that had anomalously high errors in the waveform). 
The median waveform difference (which is similar to the mismatch) between resolutions over
the simulations in the catalog is $4\times10^{-4}$. 
The simulations have a median of 
\mediannumberoforbits orbits, while the longest simulation has 
\largestnumberoforbits orbits. We have corrected each 
waveform in the catalog to be in the binary's center-of-mass 
frame and exhibit gravitational-wave memory.  We estimate the total CPU cost of
all simulations in the catalog to be \simscharged core-hours.
We find that using spectral methods for binary black hole simulations is over
1,000 times more efficient than much shorter finite-difference simulations of
comparable accuracy.
The full catalog is publicly available through the \software{sxs}
Python package and at
\url{https://data.black-holes.org}\,.
\end{abstract}

\setlength{\mathindent}{\savedmathindent}

\setcounter{tocdepth}{2}
\tableofcontents

\newpage

\renewcommand{\sectionmark}[1]{\markboth{{\scshape\thesection.\ #1}}{} }

\vspace{-1em}

\section{Introduction}

Since the
discovery of gravitational waves (GWs) from binary black holes (BBHs) in
2015~\cite{Abbott:2016blz, TheLIGOScientific:2016qqj, TheLIGOScientific:2016agk,
  TheLIGOScientific:2016wfe}, the Laser Interferometer Gravitational-Wave Observatory (LIGO)~\cite{TheLIGOScientific:2014jea} and
Virgo~\cite{TheVirgo:2014hva} have observed GWs from
the inspiral, merger, and ringdown of
dozens of BBHs~\cite{LIGOScientific:2018mvr, LIGOScientific:2020ibl, KAGRA:2021duu, KAGRA:2021vkt,
  Nitz:2021zwj}. Inferring the properties of the black holes that emitted these
waves has revealed a population of stellar-mass BBHs in a
variety of different configurations. Understanding these black-hole properties, 
such as their masses and spins, constrains stellar evolution models and enables tests of general relativity.

Achieving this understanding requires comparison of the observed GW strain with highly accurate models of the observed BBHs and the GWs they emit. Because of the nonlinear nature of general relativity,
analytic approximations like post-Newtonian~\cite{LoDr.17, Einstein:1938yz,
Damour:1985, DaDe.86, Blanchet:2013haa, Rothstein:2014sra, Porto:2016pyg,
Schafer:2018jfw, Levi:2018nxp, Futamase:2007zz, Blanchet:2009ggi,
Schaefer:2009ehj, Foffa:2013qca} and post-Minkowskian~\cite{Westpfahl:1979gu,
Westpfahl:1980mk} break down near the merger. As a result, waveform models rely
on numerical-relativity simulations to provide a ground truth to build on.
Following the 2005 breakthroughs in numerical-relativity calculation of the
inspiral, merger, and ringdown of two black holes~\cite{Pretorius:2005gq,
  Campanelli:2005dd, Baker:2005vv}, several research groups have used different
codes to create catalogs of numerical gravitational waveforms for BBHs in a variety of configurations. These include the
NINJA~\cite{Aylott:2009tn, Ajith:2012az}, NRAR~\cite{Hinder:2013oqa},
MAYA~\cite{Jani:2016wkt, Ferguson:2023vta}, RIT~\cite{Healy:2017psd,
  Healy:2019jyf, Healy:2020vre, Healy:2022wdn}, NCSA~\cite{Huerta:2019oxn},
BAM~\cite{Hamilton:2023qkv}, and GR-Athena++\cite{Rashti:2024yoc} catalogs.

Since
numerical relativity simulations are both computationally expensive and
performed for specific parameter values (e.g., mass ratio and spins), they are
typically not used in data analysis directly (although parameter estimation can
directly use numerical relativity waveforms~\cite{Lange:2017wki}). Instead, GW
model makers typically use numerical-relativity waveforms to calibrate,
validate, and build their models; specific examples include calibrating
Effective-One-Body (EOB)\cite{Buonanno:1998gg, Buonanno:2000ef,
Buonanno:2005xu,Damour:2000we, Damour:2001tu, Buonanno:2014aza, Damour:2008yg,
Ramos-Buades:2023ehm} models, validating phenomenological
models~\cite{Khan:2015jqa, Husa:2015iqa, Hannam:2013oca, Pratten:2020ceb,
Pratten:2020fqn, Ajith:2007qp, Ajith:2007kx, Ajith:2009bn, Santamaria:2010yb,
London:2017bcn, Khan:2018fmp, Khan:2019kot, Dietrich:2018nrt, Dietrich:2019nrt,
Thompson:2020nei, Garcia-Quiros:2020qpx, Garcia-Quiros:2020qlt}, and enabling the
construction of surrogate models\cite{Blackman:2017dfb, Blackman:2017pcm,
Varma:2019csw, Blackman:2015pia, Varma:2018mmi} that directly interpolate
between numerical-relativity waveforms across parameter space.

Future gravitational wave detectors, like Cosmic Explorer
(CE)~\cite{Reitze:2019iox, Evans:2023euw}, Einstein Telescope
(ET)~\cite{Punturo:2010zz, ET:2019dnz, Abac:2025saz}, the Laser Interferometer
Space Antenna (LISA)~\cite{2017arXiv170200786A}, TianQin~\cite{TianQin:2015yph},
Taiji~\cite{2021CmPhy...4...34T}, DECIGO~\cite{Kawamura:2006up}, and the Lunar
Gravitational-Wave Antenna (LGWA)~\cite{Ajith:2024mie}, will detect
GWs from BBHs much more often and with much higher
precision. Next-generation detectors on Earth will require waveform models that
are approximately one order of magnitude better than
today's~\cite{Purrer:2019jcp}, while detectors in space pose an even greater
challenge, possibly requiring several orders of magnitude increases in
accuracy~\cite{LISAConsortiumWaveformWorkingGroup:2023arg}. The challenge is
compounded in that numerical waveforms must also be both longer and more
accurate. Numerical relativity alone is too computationally expensive to cover
the needed frequency bands, so waveform models will rely on hybridization
procedures~\cite{Santamaria:2010yb, MacDonald:2011ne, Boyle:2011dy,
MacDonald:2012mp, Varma:2018mmi, Sadiq:2020hti, Mitman:2021xkq, Mitman:2022kwt,
Sun:2024kmv} that combine post-Newtonian and numerical relativity waveforms.
And since future detectors will see drastically more events, they might see
rarer BBHs, such as those with high eccentricity or extreme spins. Accurately modeling such configurations requires catalogs that 
better cover the extreme corners of the BBH parameter space.

\begin{figure}
    \includegraphics[width=0.98\columnwidth]{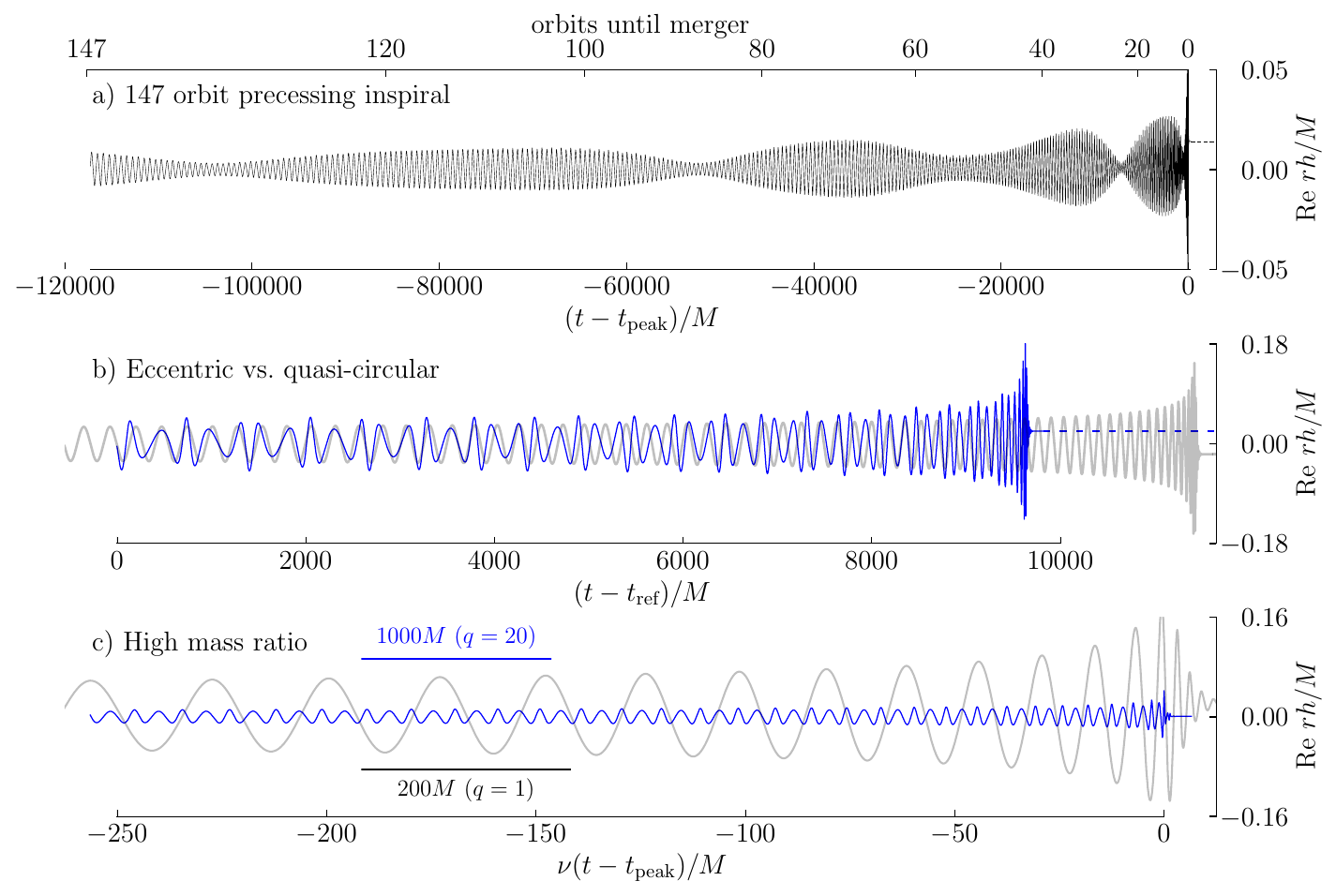}
    \caption{\label{fig:showoff}%
      An overview of some of the most extreme systems in the updated catalog.
      a)~SXS:BBH:2621, a very long, 147-orbit simulation viewed from the emission direction
      $(\theta, \phi) = (0.4\pi, 0.1\pi)$ with polarization $\psi=0.2\pi$. The non-zero value
      of the strain to the right
      of $t-t_{\mathrm{peak}}=0$ is due to gravitational wave memory
      that is included in this updated catalog. %
      b)~Blue: eccentric system SXS:BBH:2607 ($e_{\text{ref}}\approx 0.31$),
      viewed from $(\theta, \phi, \psi) = (0.3\pi, 0.5\pi, 0)$. The faint gray
      trace is a circular system SXS:BBH:1153 with the same mass
      ratio ($q=1$), time-shifted to approximately
      agree in orbit-averaged frequency at $t_{\text{ref}}$ of the eccentric
      waveform.  Note the asymmetry, higher amplitude, and
      faster merger time of the eccentric system. Gravitational wave memory is
      again indicated by the nonzero value of the strain after ringdown. %
      c)~Blue: Mass-ratio $q=20$ system SXS:BBH:2516.
      The faint gray
      trace is a $q=1$ reference system SXS:BBH:4434.
      The horizontal
      axis is scaled with the symmetric mass ratio $\nu$ so that
      the radiation-reaction timescale is the same horizontal distance
      on the plot for both waveforms.  Note the
      smaller amplitude and the much longer inspiral time of
      the high mass-ratio system. %
    }
\end{figure}

In this paper, we present a major update to the Simulating eXtreme Spacetimes
(SXS) Catalog of BBH waveforms, accessible via a python package,
\software{sxs}~\cite{sxs}. The catalog now has a total of \numberofsimulations
simulations and densely covers the parameter space, with precessing simulations
up to mass ratio $q\le8$ and dimensionless spins up to $|\chi|\le0.8$ with
near-zero eccentricity~\cite{SXSCatalogData_3.0.0}. The catalog also includes some simulations at higher mass
ratios with moderate spins, and it includes over 250 eccentric simulations.
Figure~\ref{fig:showoff} shows examples of some of the most extreme systems
present in the updated catalog, including a simulation of a precessing system
for over 147 orbits, an eccentric system with $e_{\mathrm{ref}}\approx0.31$, and
a system with mass ratio of 20.

The median waveform difference (which is similar to the mismatch, see Sec.~\ref{sec:waveform_comparison}) between resolutions over
all simulations is $4\times10^{-4}$, with a median of
\mediannumberoforbits orbits, while the longest simulation is
\largestnumberoforbits orbits. All the waveforms in the catalog are
center-of-mass and gravitational-wave-memory corrected~\cite{Woodford:2019tlo,
Mitman:2020bjf}. We have also deprecated and rerun
simulations created with a much older version of \software{SpEC} and some
simulations with anomalously high errors in the waveform. 

We estimate the total CPU cost of all the simulations in the catalog to be
only \simscharged
core-hours. Using spectral methods for long, precessing
BBH inspiral-merger-ringdown simulations is over 1,000 times more efficient than
using finite-difference methods for a few orbits of non-spinning BBHs at
comparable accuracies; see, e.g.,~\cite{Rashti:2024yoc}. This performance gap is
so large that even GPU-based finite-difference codes have yet to prove
competitive with CPU-based codes using spectral methods.

The rest of this paper is organized as
follows.
In Sec.~\ref{sec:catalog_overview}, we present an
overview of the catalog, including our catalog's coverage in terms of
parameter space and length, as well as estimates of the accuracy of the catalog's
waveforms.
In Sec.~\ref{sec:SpECsum}, we summarize the methods that we use in the
Spectral Einstein Code (\software{SpEC}), highlighting improvements since our 2019 catalog
update~\cite{Boyle:2019kee}.
In Sec.~\ref{sec:data_management}, we discuss details of how we manage the
catalog data. We briefly conclude in Sec.~\ref{sec:conclusion}. In
the appendices, we document our waveform metadata format in
Appendix~\ref{sec:metadata_fields}, describe the algorithm for determining which
simulations supersede deprecated ones in
Appendix~\ref{sec:superseded_simulations}, document our current waveform
format, including recent improvements, in Appendix~\ref{sec:waveform_format},
and list a few individual simulations with large errors
in Appendix~\ref{sec:simul-with-large}.

\section{Catalog overview}
\label{sec:catalog_overview}

\subsection{Available data}
\label{sec:available_data}

The catalog currently consists of \numberofsimulations simulations.  Each
simulation has a unique identifier of the form
\texttt{SXS:BBH:1234}.\footnote{These numbers are not necessarily consecutive.
Also note that in this paper, we only discuss BBH systems; we will
discuss black-hole--neutron-star (\texttt{BHNS}) and neutron-star--neutron-star
(\texttt{NSNS}) systems in the SXS catalog in future work.}  Each simulation typically
includes multiple otherwise-identical runs at different spatial resolutions,
which are denoted by \texttt{Lev}$N$.  These resolution numbers do not
necessarily have a consistent meaning across the catalog, but \emph{for a given
simulation} greater numbers represent higher resolution. See
Sec.~\ref{sec:waveform_comparison} for how resolution numbers are defined
for the newest simulations. The
catalog currently contains \totalnumberoflevs \texttt{Lev}s, which gives an
average of \averagenumberoflevs per simulation.

In the catalog, we include metadata (see Appendix~\ref{sec:metadata_fields} for
details) for each resolution, which provides information like masses and spins,
computed both for the initial data and at a \emph{reference time} after initial
transients have decayed away. We describe how we define the reference time in
Sec.~\ref{sec:reference-time-algorithm}.

We also provide apparent-horizon data, such as trajectories and spins as
functions of time, for both the inspiral and ringdown in the same format as in
earlier releases of the SXS catalog.  See Ref.~\cite{Boyle:2019kee} for details.
At time $t=0$, in the coordinates in which we measure trajectories and spins,
the larger black hole (BH) is on the positive $x$ axis, the smaller BH is on the negative $x$
axis, and the orbital angular momentum is in the positive $z$ direction.
However, at the reference time the BHs have moved from the positions they had at
$t=0$, so, e.g., a reference-time spin in the $x$ direction does not lie along the
line segment separating the black holes.

We extract gravitational waveforms on a series of spheres surrounding the
binary. We \emph{directly} extract the strain $h$ using Sarbach and Tiglio's
formulation~\cite{Sarbach:2001qq} of the Regge-Wheeler and Zerilli
equations~\cite{Regge:1957td, Zerilli:1970se}, with implementation details
described in~\cite{Rinne:2008vn,Boyle:2019kee}. We separately extract the
complex Weyl component $\Psi_4$, as explained in detail in
Ref.~\cite{Boyle:2019kee}.  We emphasize that these really are separate
quantities; the extraction of $h$ does not involve integrating $\Psi_4$; we
sometimes compare two time derivatives of $h$ versus $\Psi_4$ as one of our
error estimates (e.g.,
Fig.~\ref{fig:Extrapolation_order_and_Psi4_constraint_comparison}). Note that
sign conventions for quantities like $h$ and $\Psi_4$ vary in the literature;
see Appendix C. of Ref.~\cite{Boyle:2019kee} for a detailed discussion of our
sign conventions. We compute the quantities $h$ and $\Psi_4$ on multiple
coordinate spheres, typically 24 of them, that are spaced roughly uniformly in
inverse radius and extend most of the way to the outer boundary. We then
extrapolate the waveforms at these finite-radius locations to future null infinity $\mathscr{I}^+$.
See Ref.~\cite{Boyle:2019kee} for details of the extraction and extrapolation
procedure, except we now use the \software{scri} package~\cite{scri} for
extrapolation. In addition to the center-of-mass correction~\cite{Boyle:2015nqa,
Woodford:2019tlo}, we now also apply a memory correction described in
Sec.~\ref{sec:memory_correction}. We no longer supply finite-radius waveforms
since they are contaminated by gauge and near-field effects.

All simulations in the catalog include initial transients, including a burst of
non-astrophysical, high-frequency gravitational waves commonly called \emph{junk
radiation}. Such initial transients appear in all numerical-relativity
simulations of BBHs, because all known methods for constructing
constraint-satisfying initial data do not yield a BBH in
equilibrium emitting physically correct GWs. For a recent
discussion of junk radiation, see Sec.~I of Ref.~\cite{Pretto:2024dvx}. Unless
specifically studying these transients, users of the catalog should remove them
by discarding early times from all time-dependent catalog data.
Section~\ref{sec:reference-time-algorithm} discusses how we compute when junk
radiation is no longer present.

There is no overall mass scale for the BBH problem, so each of our
simulations can be scaled to any desired total mass.  The units of
strain waveforms in the catalog are $rh/M$ as a function of $u/M$,
where $rh$ is the product of the areal radius and the strain evaluated
at $\mathscr{I}^{+}$, $u$ is the retarded time at $\mathscr{I}^{+}$,
and $M$ is the sum of the Christodoulou masses of the BHs at
$\texttt{reference\_time}$ (see Sec.~\ref{sec:reference-time-algorithm}
for definition of $\texttt{reference\_time}$).  The dimensionful metadata and
horizon quantities are in code units, which are slightly different
from the units of the waveforms. For example, the Christodoulou masses
of the individual horizons at $\texttt{reference\_time}$, which correspond
to the quantities $\texttt{reference\_mass1}$ and $\texttt{reference\_mass2}$
in the metadata, do not sum exactly to unity (but the difference from unity
is typically in the sixth digit).  It is straightforward to rescale the
metadata or horizons to the same units as the waveforms, or to any
desired total mass.

All simulation data are available publicly~\cite{SXSCatalogData_3.0.0}, and we provide a
Python package \software{sxs}~\cite{sxs} to simplify obtaining,
managing, and analyzing the data, as described in
Sec.~\ref{sec:data_management}.

\subsection{Deprecated simulations\label{sec:deprecated simulations}}

Because of many improvements to \software{SpEC} over the years (see, e.g.,
Sec.~\ref{sec:recent-impr-spec}), newer simulations in the catalog are
generally more accurate than older ones.  Furthermore, we and others
have identified problems with some older waveforms that were not
evident until the numerical-relativity community began studying waveforms in greater detail,
including higher-order modes, ringdown spectroscopy,
gravitational-wave memory, etc.  Many of the largest problems occur in
the 174 simulations from the first SXS catalog paper
(published in 2013~\cite{Mroue:2013xna}).  For example,
Ref.~\cite{PereiraSturani2022} identified---among others---waveforms
with what they described as ``rippled-ringdown'' and
``asymptotic-ringdown'' anomalies, all of which came from this early
group.  We believe these are consistent with a lack of resolution
during ringdown that we did not correct until April 2015, well after we
produced the affected simulations.  As a result, we have deprecated many
early simulations. We have run newer simulations with the same physical
parameters but with improved techniques and higher resolutions.

We and others
have also occasionally found problems in a few
newer simulations; we have deprecated these simulations as well.
These problems were usually issues fixed by later updates to \software{SpEC}, but
some involved missing or corrupted files due to filesystem problems or
human error, noticed only after uploading the simulation to the
catalog.  We have recently put considerable effort into adding
validation stages to our pipeline that postprocesses and archives
simulations, so that we can identify more problems like these automatically as
they occur.
Some problematic simulations that we have deprecated
or fixed have also been pointed out
by Refs.~\cite{Khera:2021oca,Wang:2024iyj,Nobili:2025ydt}.
  One simulation, SXS:BBH:1131, had an error in the metadata
  caused by a bug in \software{SpEC}'s metadata writer that was
  present from September--November 2014.
  That simulation has been corrected in the current catalog,
  and remains non-deprecated.

We continue to include the deprecated simulations in the catalog, to
enable interested users to study these problematic
simulations.  But the \software{sxs} package will produce an error
when loading deprecated simulations, unless the user passes an option
to ignore the deprecation. In both cases, the \software{sxs} package
will suggest to the user a newer simulation with similar parameters,
either through the error message or a warning
if loading is forced.  See Appendix~\ref{sec:superseded_simulations}
for an overview of our algorithm for choosing the superseding
simulation.  Once a simulation has been deprecated, we no longer put
effort into keeping that simulation up to date with the rest of the
catalog.  As a result, deprecated simulations may not have consistent
metadata fields with new simulations or even among each other.
We encourage community members who make use of our simulations to report any
concerns; if necessary we can deprecate old simulations and possibly rerun them
if needed. To report a concern, open an issue at
\url{https://github.com/sxs-collaboration/sxs/issues/new?template=catalog-data-issue-template.md}.

\subsection{Parameter space\label{sec:parameter_space}}

\begin{figure}
  \includegraphics[width=\columnwidth]{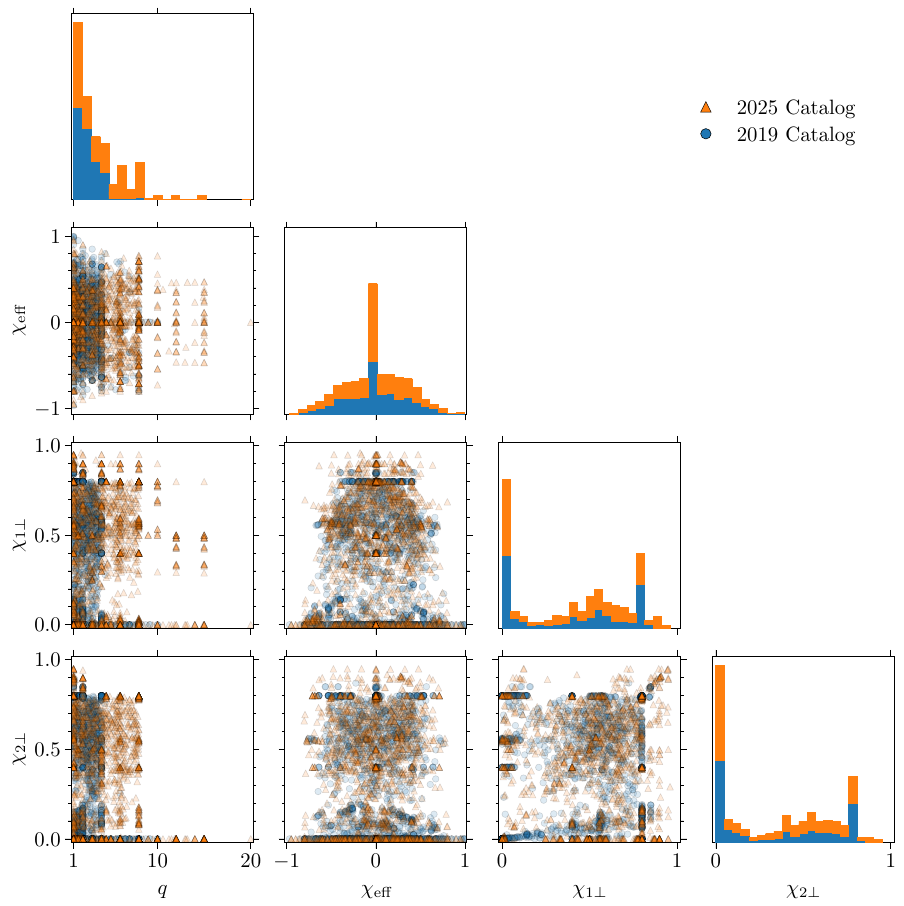}
  \caption{ \label{fig:corner_q_chieff_chi11perp_chi12perp} %
    Distribution of reference mass ratios $q$ and spins $\chi$ in the catalog.  Each panel shows a
    projection of the 7-dimensional space.  Each point is one simulation.  We
    plot the effective spin $\chi_\mathrm{eff}$ [a combination of spins that has
    a strong effect on the phasing of the gravitational waves; defined in
    Eq.~(\ref{eq:chieff})] and the magnitudes of the spins in the orbital plane.
    Blue
    circles correspond to simulations that were released as part of the 2019
    catalog, while orange diamonds correspond to simulations new
    in this release. Darker
    regions are more densely covered.
    Deprecated simulations are omitted.
  }
\end{figure}

Since the release of the 2019 catalog~\cite{Boyle:2019kee}, most of
our effort for expanding the catalog has focused on higher mass ratios
and spins.
The BBH parameter space is quantified by the
Christodoulou masses of the individual BHs $m_1$ and $m_2$ with mass ratio
$q=m_1/m_2\ge1$, dimensionless spins $\vec{\chi}_1$ and $\vec{\chi}_2$,
eccentricity $e$, and mean anomaly $\ell$ (see Sec.~\ref{sec:eccentricity}).
The dimensionless spins have magnitudes $|\vec{\chi}_{i}|\le1$; see
Sec.~2.2 of~\cite{Boyle:2019kee} for the definitions of masses and spins.
Figure~\ref{fig:corner_q_chieff_chi11perp_chi12perp} shows
mass ratios of all systems in the catalog, along
with projections of the spins into the orbital plane
\begin{equation}
  \chi_{1\perp} = \left| \vec{\chi}_1 \times \hat{L}\right|
  \qquad
  \chi_{2\perp} = \left| \vec{\chi}_2 \times \hat{L} \right|,
\end{equation}
and the effective spin~\cite{Ajith:2009bn, Santamaria:2010yb,
Hannam:2013oca}
\begin{equation}
  \label{eq:chieff}
  \chi_{\mathrm{eff}}
  \equiv \frac{(m_1\vec{\chi}_1 + m_2 \vec \chi_2)\cdot \hat{L}} {m_1+m_2}
  = \frac {m_1 \chi_{1\parallel} + m_2 \chi_{2\parallel}}{m_1+m_2},
\end{equation}
where $\hat{L}$ is the direction of the instantaneous Newtonian
orbital angular momentum. We extract all quantities at $\texttt{reference\_time}$.
In particular, we now have dense coverage in mass ratios up to $q=8$, including
large spins and significant precession.  We have also performed over 100
simulations between $q=8$ and $q=20$---most with essentially no spin on the
smaller black hole, but significant (usually precessing) spin on the larger black hole.

\begin{figure}
  \includegraphics[width=\columnwidth]{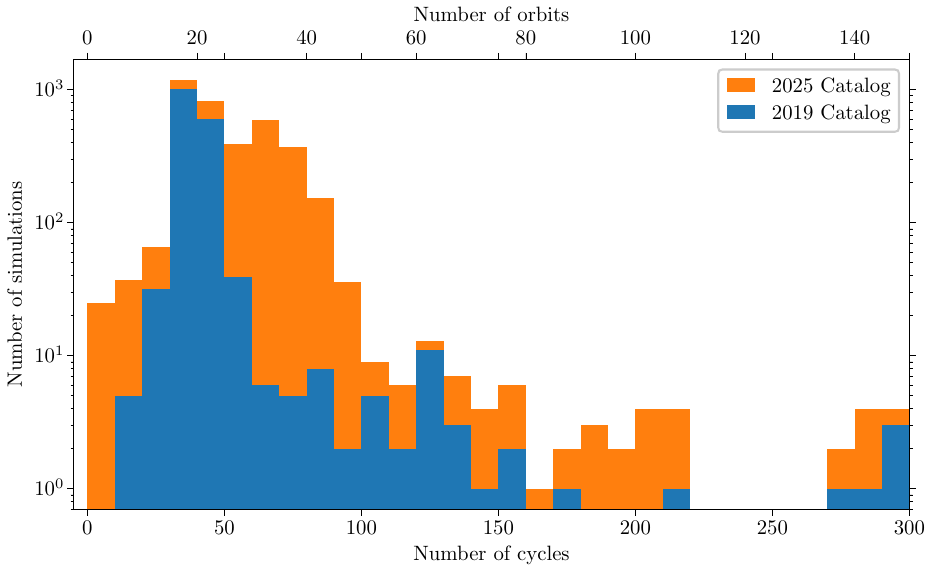}
  \caption{ \label{fig:histogram_n_cycles} %
  The number of orbits (top axis) and number of cycles (bottom axis) of the
  $\ell = m = 2$ GWs from the start of the simulation until the
  formation of a common apparent horizon for the simulations in the
  catalog, as
  determined by the
  coordinate trajectories of the black holes.  Bin edges are multiples of 5
  orbits and 10 cycles.
  Deprecated simulations are omitted.
}
\end{figure}

\begin{figure}
        \includegraphics[width=\columnwidth]{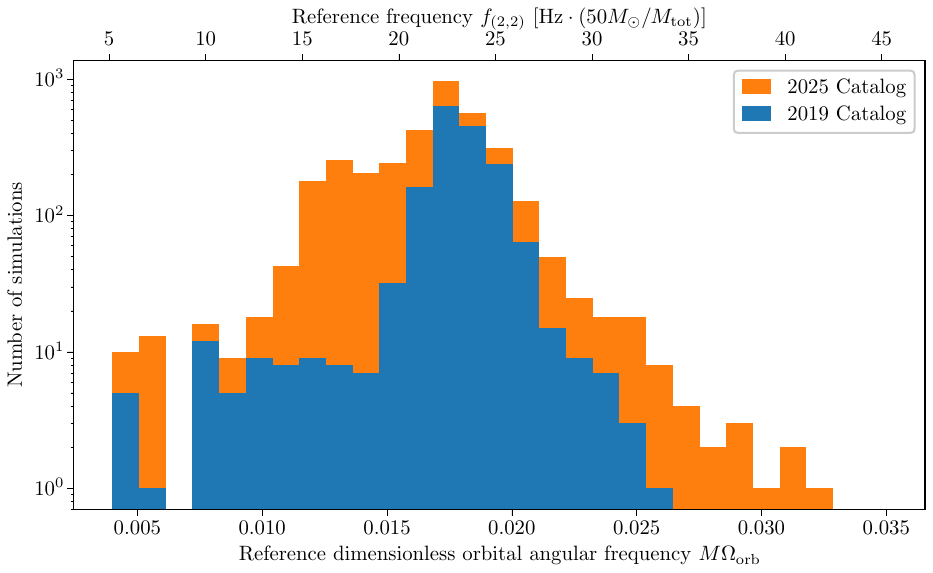}
        \caption{ \label{fig:histogram_min_frequency} %
          The reference dimensionless orbital angular frequency
          $M\Omega_{\mathrm{orb}}$ for the simulations in the
          catalog. The top axis is the frequency of the $(2,2)$ mode
          at the reference time for a binary with a total mass of
          $50\mathrm{M_{\odot}}$. Only simulations with a reference
          eccentricity $<10^{-2}$ are shown.
          Deprecated simulations are omitted.
        }
\end{figure}

We have continued to increase the number of cycles of our simulations,
even for computationally challenging cases with high mass ratios and
spins.  A new subset of about 1,000 precessing simulations with
$q\leq 8$ has approximately 33 orbits per simulation, corresponding to
$\approx66$ GW cycles.  Figure~\ref{fig:histogram_n_cycles} shows, for
all simulations in the catalog, a histogram of the number of
simulations binned by the number of GW cycles. Compared to the
previous catalog, the figure highlights an increase in the number of
simulations that have 50--70 GW cycles.\footnote{We deprecated the longest
  simulation in the 2019 catalog, SXS:BBH:1110, because the waveform was
  contaminated with a center-of-mass acceleration as reported
  in Ref.~\cite{Szilagyi:2015rwa}.}
There are also a number of new simulations with very few orbits that
represent nearly-head-on collisions and scattering scenarios
that are now in the catalog.
We have produced a small
number of simulations with even more cycles, but we have primarily
focused on improved parameter-space coverage rather than on producing
longer waveforms.  This is for several reasons: First, 50--70 GW
cycles is sufficient for most GW applications today, especially for systems
with higher total masses where LIGO-Virgo-KAGRA (LVK) detectors are
not sensitive enough at low frequencies to detect a longer inspiral.
Second, for low-mass systems that can have many more
gravitational-wave cycles in band than would be feasible for us to
simulate, alternative approaches (e.g., hybridizing numerical
relativity and post-Newtonian models) can yield waveforms with
sufficient length, although it remains unclear how long the numerical
relativity waveforms must be for accurate hybridization, especially
for precessing systems~\cite{Sun:2024kmv}.  Finally, as demonstrated
in~\cite{Mitman:2025tmj} and discussed in
Sec.~\ref{sec:waveform_comparison} below, the mismatch and waveform
difference increase as $\sim t^2$, meaning longer simulations require
significantly higher resolution to maintain reasonable phase
errors. Nonetheless, in Fig.~\ref{fig:histogram_min_frequency} we plot the
reference dimensionless orbital angular frequency $M\Omega_{\mathrm{orb}}$ for
our simulations to show what our low frequencies are. The top axis shows the
dimensionful frequency of the $(2,2)$ mode at the reference time for a system with a total mass of
$50M_{\odot}$.
Only simulations with reference eccentricities $<10^{-2}$ are shown
in Fig.~\ref{fig:histogram_min_frequency}, since
for eccentric cases the instantaneous reference frequency
$\Omega_{\mathrm{orb}}$ does not necessarily correspond to the
lowest frequency in the waveform.

\begin{figure}
  \includegraphics[width=\columnwidth]{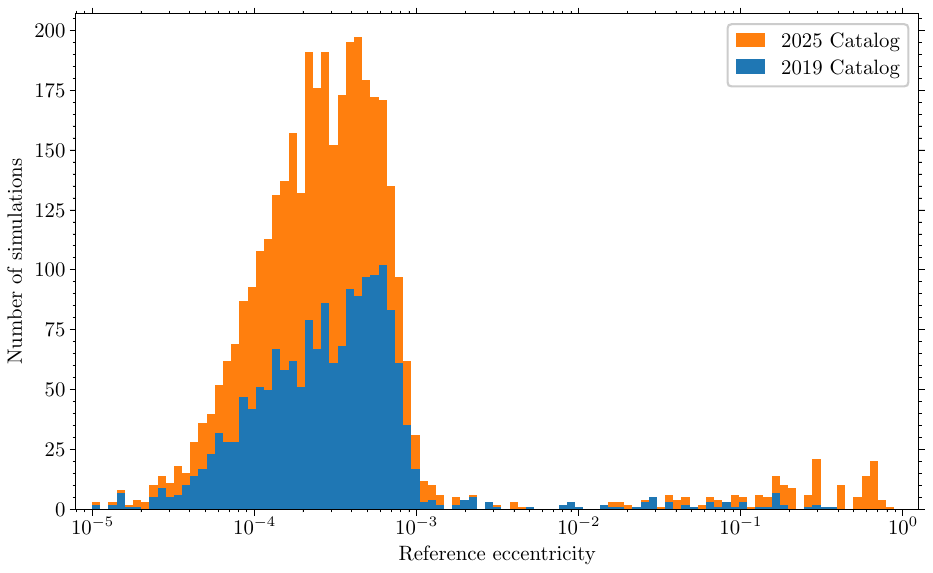}
  \caption{\label{fig:histogram_eccentricity}
  The number of simulations at different reference eccentricities $e_{\mathrm{ref}}$ in the
  catalog. The main population shows simulations using
  eccentricity reduction, while we have also completed several campaigns targeted at high $e_{\mathrm{ref}}$,
  yielding the tail.
  Deprecated simulations are omitted.
  }
\end{figure}

Figure~\ref{fig:histogram_eccentricity} shows a histogram illustrating
the number of simulations with different orbital eccentricities.
In Sec.~\ref{sec:eccentricity}, we
describe the algorithm we use to measure the eccentricities,
which is based on the black hole
trajectories, not the waveforms. Because the method uses trajectories,
and because of the lack of a unique definition of eccentricity in
general relativity (though efforts have been made to relate and
understand different definitions, e.g.,~\cite{Shaikh:2023ypz}), we
recommend using the ``reference eccentricity'' value in the simulation
metadata, as computed according to Sec.~\ref{sec:eccentricity}, only
as a rough estimate. For more detailed analyses that depend on precise
values of eccentricity, users should choose a definition of
eccentricity and consistently measure it from the gravitational
waveform.  In general, we intend simulations with reference
eccentricity below $10^{-3}$ in
Figure~\ref{fig:histogram_eccentricity} to be quasi-circular; for
these, we have used an iterative eccentricity-reduction scheme. In
contrast, we intend simulations with reference eccentricity above
$10^{-3}$ to represent eccentric systems.
Figure~\ref{fig:histogram_eccentricity} shows that the majority of
waveforms in the catalog are quasi-circular,
as in our 2019 catalog.
However, we are continuing our efforts to extend the
catalog to more eccentric and precessing systems~\cite{Islam:2021mha,
  Ramos-Buades:2022lgf}. We will include more eccentric waveforms in
future updates and releases of the catalog.

In summary, since 2019, we have nearly doubled the size of the catalog, with most new
simulations having precessing spins with mass ratios $q>4$. We also include a significant
number of eccentric systems, as well as some
simulations of hyperbolic encounters and scattering scenarios.

\subsection{Waveform comparison}
\label{sec:waveform_comparison}

For measuring the accuracy of our waveforms, we
compare the strain computed by (at least) two different
resolutions for each simulation.
For the newest simulations, we define different resolutions
such that the target relative truncation error of the metric and its
derivatives
in the wavezone
for \texttt{Lev}$k$ is given by $2.17\times10^{-4}\times4^{-k}$, and
near the black holes it is approximately two orders of magnitude
smaller. We set the projected constraint error (see \cite{Szilagyi:2014fna})
tolerance to be four orders of magnitude smaller than the
wavezone truncation error tolerance.  Note that resolution numbers
should be directly compared only for the same simulation.
Resolution numbers
for different simulations do not necessarily correspond to the same
final errors, since different simulations (even newer ones) sometimes
vary in choices of initial data or gauge (see Sec.~\ref{sec:SpECsum}), as well as masses, spins, and number of orbits.

When presenting errors, we consider two quantities. The first
is the waveform
difference\footnote{We choose this form so that the difference and
mismatch agree in the limit of a flat power
		spectral density and infinite numerical resolution.}
\begin{equation}
	\label{eq:waveform difference}
	\Delta(h_1,h_2)=\frac{1}{2}\frac{\|h_1-h_2\|^2}{\sqrt{\|h_1\|^2 \|h_2\|^2}},
\end{equation}
where
\begin{equation}
	\label{eq:waveform norm}
	||h_{1}||=\sqrt{\int_{t_{1}}^{t_{2}}\int_{S^{2}} |h_{1}|^{2}\,d\Omega\,dt}.
\end{equation}
The second is the waveform mismatch averaged over the two-sphere (which we hereafter refer to as the averaged mismatch)
\begin{equation}
	\label{eq:waveform mismatch}
	\overline{\mathcal{M}}(h_{1},h_{2})=1-\frac{\langle h_{1},h_{2}\rangle}{\sqrt{\langle h_{1},h_{1}\rangle\langle h_{2},h_{2}\rangle}},
\end{equation}
where the inner product $\langle h_{1},h_{2}\rangle$ is
\begin{align}
	\label{eq:inner product}
	\langle h_{1},h_{2}\rangle=\mathrm{Re}\left[\int_{f_{1}}^{\infty}\int_{S^{2}} \frac{\tilde{h}_{1}(f)\tilde{h}_{2}(f)^{*}}{S_{n}(f)}\,d\Omega\,df\right].
\end{align}
In Eq.~\eqref{eq:waveform norm}, $t_{1}$ is the relaxation time defined in
Sec. \ref{sec:reference-time-algorithm}, and
$t_{2}=t_\mathrm{merger}+0.6(t_\mathrm{end}-t_\mathrm{merger})$,
where $t_{\mathrm{merger}}$ is the time at which the strain's $L^{2}$ norm over the two-sphere achieves its maximum value and $t_\mathrm{end}$ is the end time of the simulation.  For those simulations without
mergers (e.g., hyperbolic encounters),
we choose $t_{2}=t_\mathrm{end}-100M$.
In Eq.~\eqref{eq:inner product}, $\tilde{h}_{1}(f)$ is the Fourier transform of $h_{1}(t)$, $\phantom{}^{*}$ represents complex conjugation, $f_{1}$ is twice the orbital frequency at the relaxation time, and $S_{n}(f)$ is the power spectral density of a GW detector. Note that $\overline{\mathcal{M}}$
differs from the mismatch used in
Ref.~\cite{Boyle:2019kee} and in typical LVK analyses, which instead compute mismatches at various points on the two-sphere, minimizing the mismatch over time and phase shifts at each point on the sky independently, and then report the largest mismatch. The averaged mismatch in
Eq.~\eqref{eq:waveform mismatch}, however, corresponds to an average mismatch over the whole
two-sphere since no point evaluation is performed. We use this new quantity
because it
better quantifies our waveform differences
by including correlations between different points on the two-sphere that are
important for getting spherical-harmonic modes of the waveform correct for
modeling purposes.
  All of the code to compute mismatches and
  reproduce the subsequent analysis is
available via the \texttt{sxs.simulations.analyze.analyze\_simulation} function.

It is customary to consider two waveforms to be equal if they differ
only by an overall phase or time shift, since
the two waveforms represent the same physics. Similarly, two
waveforms should also be considered equal if they differ only by an $SO(3)$ rotation, Lorentz boost, or supertranslation. There are a few ways in which one
can incorporate these frame freedoms into a comparison between
waveforms at different resolutions. First, one could fix the frame
freedom of the waveform at each resolution independently, e.g., make
$t=0$ correspond to the time at which the strain's $L^{2}$ norm
achieves its maximum value and fix the other Bondi-van der
Burg-Metzner-Sachs (BMS) freedoms in a well-prescribed way; we call
this option the ``independent alignment'' method.  Second, one could
find the optimal BMS transformation that makes the waveform from one
resolution best agree with the waveform from the other resolution; we
call this the ``minimal difference'' method since the optimization yields the
smallest difference.

For the majority of our convergence tests, we use the independent
alignment method, as it fixes the coordinate freedom of each
resolution independently and will still converge in the limit of
infinite resolution. This method is appropriate for waveform models
that claim to be in some well-defined coordinate frame, because
failures of the model to accurately fix that frame will translate into
differences between waveforms.
However, because no optimization over the coordinate
transformations is performed, the difference produced by the independent
alignment method will be larger than the difference resulting from
methods that do optimize over these transformations.  Thus, for
illustrative purposes, for some cases we will compare with the minimal
difference method, for which we perform a four-dimensional optimization
over a time translation and $SO(3)$ rotation to minimize the waveform
difference (see Eq.~\eqref{eq:waveform difference}). Note that we do not optimize over Lorentz boosts
and supertranslations, since these optimizations can be very expensive
and tend to be less important than optimizing the time and rotation freedoms, which are more directly related to the phase error of the simulation. Nonetheless, methods for efficiently optimizing over all of the BMS freedoms are in development and will be performed for future catalog analyses.

For the independent alignment method, we fix the frame freedom as
follows. First, we perform a time translation so that $t=0$ of each waveform
corresponds to the merger, which here we define as the time at which the
strain's $L^{2}$ norm over the two-sphere achieves its maximum value. Then, we find the rotation
that aligns the angular velocity with the positive
$z$ axis~\cite{Boyle:2013nka}, which makes the phase of the $(2,2)$ mode
zero and the real part of the $(2,1)$ mode positive. We do this at the
time that is 10\% the length of inspiral (not including the times before the
relaxation time) before $t=0$. For example, if the post-relaxation time inspiral
is 5,000$M$ long, then we fix the frame at $-500M$.

\begin{figure}
  \includegraphics[width=\columnwidth]{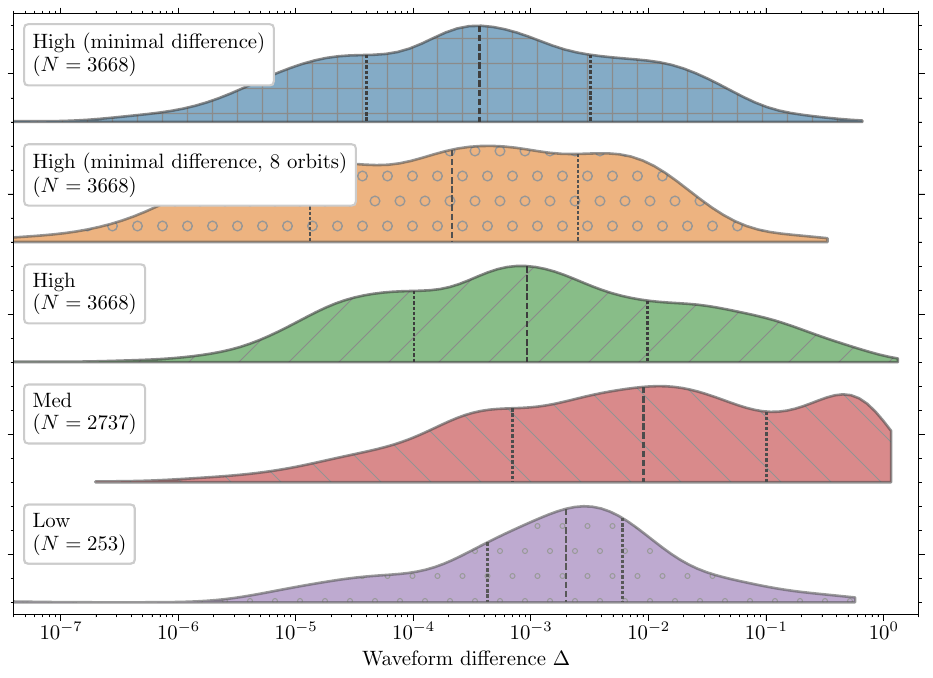}
  \caption{ \label{fig:Lev_comparison} %
    Waveform difference (see Eq.~\eqref{eq:waveform difference}) between strain waveforms for different simulation resolutions. Blue
    corresponds to the difference between the two highest resolutions available,
    using the minimal difference alignment method.
    Orange is identical to blue, but with
    the waveform truncated so that it starts 8 orbits before merger.
    Green is identical to blue, but with the independent alignment method.
    Red and purple are identical to green, but for the next
    two highest pairs of resolutions. Vertical lines show the quartiles of each
    distribution. The number of simulations in each distribution is shown in the legend. While some simulations have differences as low as $10^{-10}$,
    we truncate the horizontal axis to make the figure more readable.
  }
\end{figure}

\begin{figure}
	\includegraphics[width=\columnwidth]{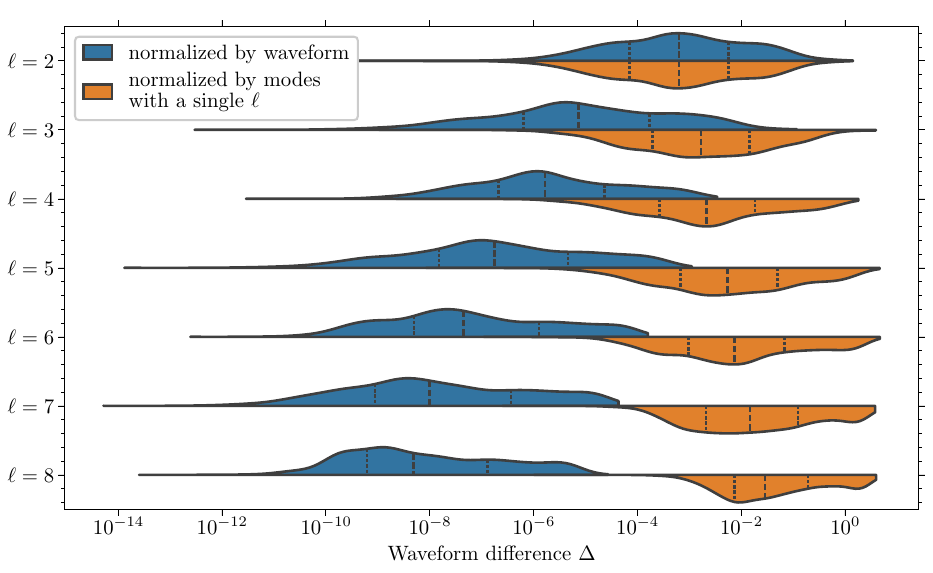}
	\caption{ \label{fig:Lev_comparison_per_ell} %
		Waveform difference (per $\ell$) (see Eq.~\eqref{eq:waveform difference}) between strain waveforms for the two 
		highest-resolution simulations. In blue we show waveform differences normalized by the full
		waveform, while in orange we show differences normalized by only the
		$\ell'$ harmonics with $\ell'=\ell$. The waveforms used for these errors have been aligned using the minimal difference method. Vertical lines show the quartiles of each distribution.
}
\end{figure}

In Fig.~\ref{fig:Lev_comparison} we show the waveform difference between the resolutions
available for each simulation. All waveform differences and mismatches
used to produce Figs.~\ref{fig:Lev_comparison}--\ref{fig:Lev_comparison_ASDs_vs_f_low}
are available in the supplementary material~\cite{JSONSupplement}.
Note that we have a handful of simulations with a single resolution, which do
not contribute to this figure. The blue (green) curve in
Fig.~\ref{fig:Lev_comparison} shows the difference
between the two highest resolutions when using the minimal difference (independent
alignment) method, while the red and purple curves show the differences between
the next highest resolutions with the independent alignment method. The vertical
dashed lines show the quartiles of each distribution. As can be seen by
comparing the blue and green curves, the minimal difference method improves the
median difference by a factor of two and lessens the high-difference tail, which
results from some of the more precessing systems requiring a much finer tuning
of the rotation transformation. The reason why the purple distribution appears to
have smaller differences than the red distribution is simply because there are
fewer data points for the purple distribution, since we have fewer simulations that have
four separate resolutions. Also, those simulations with four separate
resolutions tend to be in less extreme regions of parameter space, where
computational cost is lower and where it is easier to obtain higher accuracy.
This figure shows that the relative error in our
waveforms over the entire catalog, on average, is $\mathcal{O}(1\%)$. In terms
of mismatches (see also Fig.~\ref{fig:Lev_comparison_ASDs_vs_f_low}), this
median waveform difference corresponds to a averaged mismatch of
$\mathcal{O}(10^{-4})$. 

An important note regarding these waveform differences, however, is that they are highly dependent on
the length of the simulation. In particular, in Ref.~\cite{Mitman:2025tmj} it
was shown that the mismatch (and therefore also the waveform difference)
between numerical relativity waveforms from
different resolutions, whose relative error is dominated by some phase
difference $\delta\phi$, tends to scale as
\begin{align}
  \label{eq:mismatch phase scaling}
  \overline{\mathcal{M}}\left(h,e^{i\,\delta\phi\,t}h\right)\sim\delta\phi^{2}|t_{2}-t_{1}|^{2}.
\end{align}
Therefore, while the differences in the blue curve Fig.~\ref{fig:Lev_comparison} may seem high
compared to other catalogs', this is mainly because our simulations tend to be
very long (see Fig.~\ref{fig:histogram_n_cycles} for an overview of our
lengths). In particular, any numerical relativity waveform dominated by phase error
can be made to have a smaller difference simply by 
shortening it by removing the beginning of the waveform. We demonstrate this with the orange curve,
for which we truncate our simulations to be 8 orbits before computing waveform differences
using the minimal difference method. In the most extreme case, by truncating one of our
simulations to only be 8 orbits, we can improve the waveform difference between
resolutions by four orders of magnitude. The remaining high tail in the orange curve
is due to a few highly precessing systems that are underresolved, so that
the two resolutions end up describing systems that are slightly physically
different (e.g., different spin directions).

For building waveform models, the errors in individual spherical-harmonic modes of the strain are often of interest.
In Fig.~\ref{fig:Lev_comparison_per_ell} we show the
waveform difference between the
two highest resolutions when restricted to spherical-harmonic modes
with a certain $\ell$. In blue we
show the difference normalized by the norm of the entire waveform while in
orange we show the difference normalized by the norm of only the modes with
that particular value of $\ell$. All of the errors are calculated from waveforms aligned with the minimal difference method. As may be expected because of their higher frequency and more complicated sourcing,
the larger $\ell$ modes exhibit larger errors relative to their
amplitudes.  However, the larger $\ell$ modes contribute much less to the
  overall strain, so the errors in these modes have a small contribution to
  the overall waveform. Comparing Fig.~\ref{fig:Lev_comparison_per_ell} to the blue distribution in Fig.~\ref{fig:Lev_comparison} makes it apparent that our dominant source of error in the overall waveform is still from the $\ell=2$ modes.

\begin{figure}
  \includegraphics[width=\columnwidth]{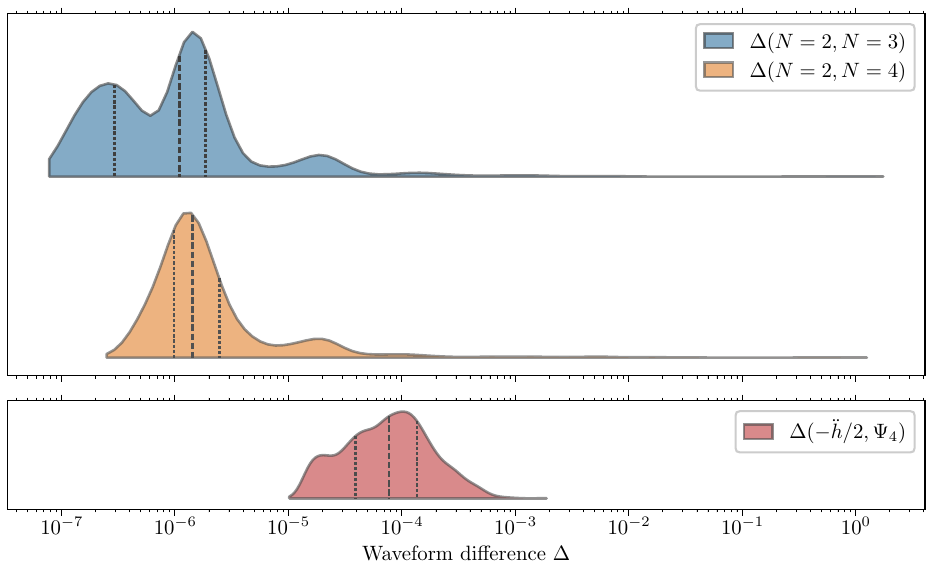}
  \caption{\label{fig:Extrapolation_order_and_Psi4_constraint_comparison} %
    Top panel: waveform difference (see Eq.~\eqref{eq:waveform difference})
    between different extrapolation orders. Bottom panel: Difference
    between $\Psi_{4}$ and $-\ddot{h}/2$, which are computed independently
    by different methods but which should agree.  Differences are taken
    for the highest resolution for each
    simulation. Waveforms are aligned using the independent alignment
    method. Vertical lines show the quartiles of each distribution.
  }
\end{figure}

The procedure~\cite{Boyle:2019kee} used
to extrapolate a series of finite-radius waveforms
to produce a waveform at $\mathscr{I}^+$ uses a polynomial fit
of order $N$ to extrapolate in $1/r$.
Varying this order $N$ can be used to quantify the
error of the extrapolation procedure.
In Fig.~\ref{fig:Extrapolation_order_and_Psi4_constraint_comparison}, we show the waveform difference between
different extrapolation orders for the highest resolution. The blue curve
shows $N=2$ vs.\ $N=3$ and the orange curve shows $N=2$ vs.\ $N=4$. Waveforms are aligned using the independent alignment method in this figure. When comparing orders $N=2$ and $N=3$,
the differences peak near $\mathcal{O}(10^{-6})$ with a tail that extends to
$\sim1$. Overall this result demonstrates
that by this measure, errors in our extrapolation
procedure are on average unimportant when compared to the numerical truncation
error. That is, the extrapolation errors in
Fig.~\ref{fig:Extrapolation_order_and_Psi4_constraint_comparison} are typically
an order of magnitude less than the errors shown in
Fig.~\ref{fig:Lev_comparison} and are therefore not our dominant source of
waveform error. The high-$\Delta$ tail
is due to a few nearly-head-on or scattering simulations, 
which we list in Appendix~\ref{sec:simul-with-large}.
One reason for the large waveform differences is junk radiation, which causes
large differences between extrapolation orders.  The actual amount of junk
radiation is similar to that in inspiral waveforms, but in these short
simulations
we do not have the luxury of waiting longer for junk to decay before the
interesting physics occurs, so the relative contribution of the initial
transient junk radiation to the waveform difference is larger.
Another reason for the large waveform difference is that some of these runs 
have waveform-extraction radii that are very close together resulting in an
inaccurate extrapolation.  Simulations with narrow distributions of extraction
radii will be deprecated and rerun in the future.

In the bottom panel of Fig.~\ref{fig:Extrapolation_order_and_Psi4_constraint_comparison}, we also show the difference between
$-\frac{1}{2}\ddot{h}$ and $\Psi_{4}$, two quantities
that we compute independently in
our code by completely different methods, but which should agree.
The agreement between $-\frac{1}{2}\ddot{h}$ and $\Psi_{4}$
shows how well our simulations
respect the Bianchi identity for $\Psi_{4}$ on future null infinity
$\mathscr{I}^{+}$.

\begin{figure}
        \includegraphics[width=\columnwidth]{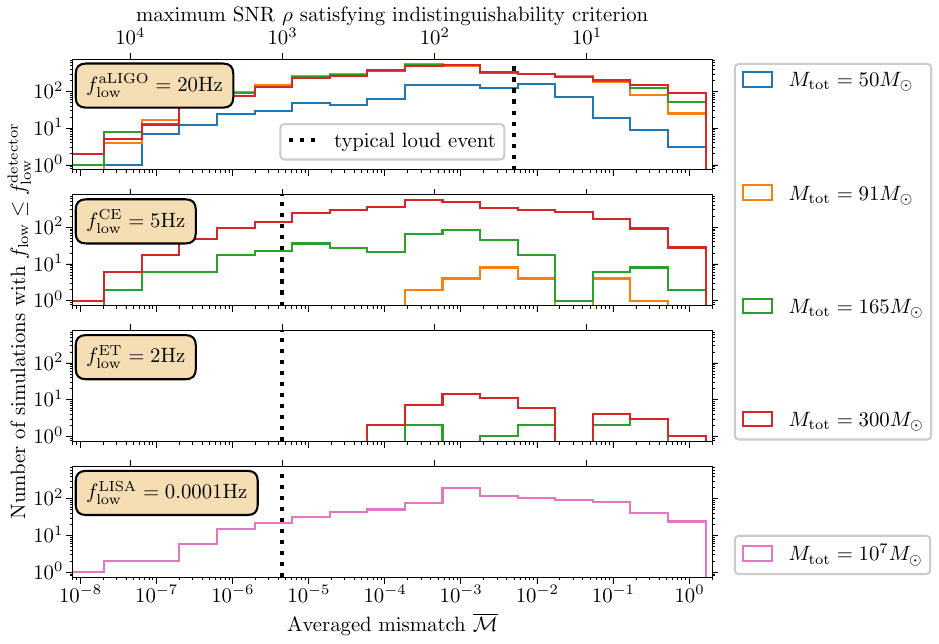}
        \caption{ \label{fig:Lev_comparison_ASDs_vs_f_low} %
          Number of simulations with
          $f_{\mathrm{low}}\leq f_{\mathrm{low}}^{\mathrm{detector}}$
          vs. averaged
          mismatch (Eq.~\eqref{eq:waveform mismatch}) between the strain
          waveforms for the two highest resolutions of that simulation.
          Here
          $f_{\mathrm{low}}$ is twice the orbital frequency at the
          simulation's relaxation time and
          $f_{\mathrm{low}}^{\mathrm{detector}}$ is the low-frequency
          cutoff of each detector.
          The top horizontal axis represents the
          signal-to-noise ratio (SNR) at which the corresponding
          mismatch will not bias data analyses, according to
          Eq.~\eqref{eq:mismatchSNR}.  All simulations to the left of
          a specific mismatch have truncation error sufficiently small
          to not bias data analyses of signals with
          SNR smaller than $\rho$ on the top axis. Different colors
          correspond to different total masses, logarithmically spaced
          from $50M_{\odot}$ to $300M_{\odot}$ for ground-based
          detectors.  The vertical dotted lines indicate the SNR of
          typical loud events observed/expected in each
          detector~\cite{KAGRA:2021vkt,Evans:2023euw,Abac:2025saz,LISAConsortiumWaveformWorkingGroup:2023arg}. Only simulations with reference eccentricities
          $<10^{-2}$ are shown.}
\end{figure}

Finally, we assess the readiness of our catalog for next-generation detectors,
specifically, LIGO Design, Cosmic Explorer
(CE), Einstein Telescope (ET), and LISA. In Fig.~\ref{fig:Lev_comparison_ASDs_vs_f_low},
we show the number of
simulations in the catalog whose lowest frequency---taken to be twice the reference orbital frequency---is
below the low-frequency
cutoff of each detector.
Only simulations with reference eccentricities $<10^{-2}$ are shown in Fig.~\ref{fig:Lev_comparison_ASDs_vs_f_low}, since
for eccentric cases the instantaneous reference frequency
does not necessarily correspond to the
lowest frequency in the waveform.
On the bottom horizontal axis of
Fig.~\ref{fig:Lev_comparison_ASDs_vs_f_low} we plot the averaged mismatch
(see Eq.~\eqref{eq:waveform mismatch})
between the two highest resolutions when using the power spectral densities (PSDs) for
the four detectors, while on the top horizontal axis we plot the
signal-to-noise ratio (SNR) at which the corresponding mismatch is sufficient to
not bias data analyses. Specifically, we use the well-known sufficient condition for two
waveform models to be indistinguishable~\cite{Flanagan:1997kp,Lindblom:2008cm,McWilliams:2010eq,Baird:2012cu,Chatziioannou:2017tdw,Toubiana:2024car},
\begin{align}
  \label{eq:mismatchSNR}
  \overline{\mathcal{M}} < \frac{D}{2\rho^2},
\end{align}
where $\overline{\mathcal{M}}$ is the averaged mismatch (see Eq.~\eqref{eq:waveform mismatch}), $D$ is the number of intrinsic parameters with $D=9$, and
$\rho$ is the SNR of the observation the models are describing. Note that Eq.~\eqref{eq:mismatchSNR} is a sufficient, but not necessary, condition; i.e., there may be some situations where an averaged mismatch larger than the right hand side of Eq.~\eqref{eq:mismatchSNR} will still not lead to large biases in data analyses~\cite{Toubiana:2024car}.
In Fig.~\ref{fig:Lev_comparison_ASDs_vs_f_low}, different colors correspond to different total masses and vertical dotted lines show SNRs of typical loud events expected for each detector. The PSDs for the four detectors come from the following: For aLIGO, we use the PSD presented in Ref.~\cite{LIGOTechnicalNote} and published in Ref.~\cite{LIGO}; for CE, we use the PSD presented in Ref.~\cite{Evans:2021gyd,Srivastava:2022slt} and published in Ref.~\cite{CE}; for ET, we use the ET-D PSD presented in Ref.~\cite{Abac:2025saz} and published in Ref.~\cite{ET}; and for LISA, we use the PSD presented in Ref.~\cite{Babak:2021mhe} and published in Ref.~\cite{LISA}.

To interpret the figure, choose a SNR for a given detector, and choose a total
mass. Then all simulations to the left of that SNR in the figure have enough
cycles to cover the frequency band of the detector, and sufficiently small
truncation error so as to not bias data analysis of a signal with that SNR, or
with any smaller SNR.  For example, for a Cosmic Explorer event with SNR
$\sim 200$, the corresponding mismatch is about $2\times 10^{-4}$, there there
are hundreds of simulations in the catalog sufficiently long and accurate for a
total mass of $165M_{\odot}$ and larger, but for that SNR there are no
simulations in the catalog sufficient for a total mass of only $91M_{\odot}$.

The figure shows that, for LIGO, which has
a much higher low-frequency cutoff, the majority of our catalog has a sufficiently small averaged mismatch
for events with $M_{\mathrm{tot}}>90M_{\odot}$. However, for lower total masses
or for future detectors with lower low-frequency cutoffs, many of our
simulations are too short to span the entire frequency band, and most of the
simulations that do span the entire frequency band may have mismatches too
large to not bias the data analysis. In particular, for ET, which has a low-frequency cut off 2Hz, we find that we have very few simulations long enough to span the frequency band for $M_{\mathrm{tot}}<300M_{\odot}$, while for LISA we have no simulations long enough for to span the frequency band for $M_{\mathrm{tot}}<10^{6}M_{\odot}$. While hybridization with post-Newtonian
waveforms may reduce the need for extremely long numerical relativity
waveforms, the exact requirements for hybridization at next-generation
accuracies have yet to be fully determined~\cite{Santamaria:2010yb,
  MacDonald:2011ne, Boyle:2011dy, MacDonald:2012mp, Varma:2018mmi,
  Sadiq:2020hti, Mitman:2021xkq, Mitman:2022kwt, Sun:2024kmv}.

\section{Numerical methods}
\label{sec:SpECsum}

All the simulations in the catalog are performed using the
Spectral Einstein Code (\software{SpEC})~\cite{SpECwebsite}.
We construct constraint-satisfying initial data using the Extended
Conformal Thin Sandwich (XCTS)~\cite{York:1998hy, Pfeiffer:2002iy}
formulation. In the XCTS formulation, the conformal 3-metric and the
trace of the extrinsic curvature, as well as their time derivatives,
can be freely chosen.  Typically we choose these quantities to be
weighted superpositions of the analytic solutions for two single black
holes in Kerr-Schild coordinates~\cite{Lovelace:2008tw}, in
time-independent horizon-penetrating harmonic
coordinates~\cite{Cook:1997qc,Ma:2021can}, in time-independent
damped harmonic coordinates~\cite{Varma:2018evz},
or in a spherical version of Kerr-Schild coordinates~\cite{Chen:2021rtb};
each of these
choices has its own advantages and drawbacks~\cite{Varma:2018sqd,Chen:2021rtb}.
A few simulations in the catalog are conformally flat and have a
vanishing trace of the extrinsic curvature (i.e., maximal
slicing)~\cite{Caudill:2006hw}. We choose the time derivatives of the
conformal metric and trace of extrinsic curvature to be zero.  We then
solve the XCTS equations on a grid with two excised regions using a
spectral elliptic solver~\cite{Pfeiffer:2002wt}. The boundary
conditions on the excision boundaries are chosen to ensure that
these boundaries are either apparent horizons~\cite{Caudill:2006hw,
  Lovelace:2008tw}, or that they are surfaces of constant expansion
with a small negative value~\cite{Varma:2018sqd} and thus are slightly
inside apparent horizons.  The outer boundary is at a radius around $\sim10^{9}M$, and the boundary conditions are derived by requiring asymptotic flatness. The solution of the XCTS equations provides
constraint-satisfying initial data on the initial slice.

The initial data depend on a set of input parameters, such as the spin
vectors and masses associated with the analytic single-black-hole
solutions used to build the conformal metric, and also the initial
coordinate positions and velocities of the black holes.  After the
constraints are solved, we compute physical parameters such as the
spin vectors and masses of the black holes, and frame quantities such
as the initial total linear momentum and center-of-mass of the binary.
These computed physical and frame quantities are not the same as the
input parameters, and cannot be computed from the input parameters
without solving the nonlinear XCTS equations.  After measuring the
physical and frame quantities, we adjust the input parameters and
re-solve the XCTS equations. This process is iterated until the black holes
have the desired masses and spins, the initial linear momentum is
zero, and the initial center-of-mass is at the
origin~\cite{Buchman:2012dw,Ossokine:2015yla}.

For most of the simulations in the SXS catalog, we carry out an
additional iteration that briefly (typically for a few orbits) evolves
the initial data, and adjusts the initial coordinate velocities to
yield a BBH with small orbital eccentricity~\cite{Pfeiffer:2007yz,
  Buonanno:2010yk, Mroue:2012kv, Habib:2024soh}, typically $e_{0} \sim
10^{-4}$.  For some simulations in the catalog, we intentionally omit
the eccentricity-reduction iteration, to obtain initial data for BBHs
with non-negligible orbital eccentricity.  For other simulations we
include an eccentricity iteration to tune the eccentricity to desired
nonzero values~\cite{Knapp:2024yww, Nee:2025zdy}. See Sec.~\ref{sec:eccentricity}
for details of eccentricity measurement.

For the evolution, we use a first-order version of the generalized
harmonic formulation~\cite{Friedrich1985, Garfinkle:2001ni,
  Pretorius:2004jg, Lindblom:2005qh} of Einstein's
equations~\cite{Gundlach:2005eh, Pretorius:2004jg, Lindblom:2005qh}.
For most simulations we choose an initial gauge that approximates a
time-independent solution in a co-rotating frame, and then we smoothly
change to damped harmonic gauge~\cite{Lindblom:2009tu,
  Choptuik:2009ww, Szilagyi:2009qz}, which works well near merger.
A few simulations start in harmonic gauge at $t=0$ and smoothly change
to damped harmonic gauge.  Some simulations start in damped harmonic
gauge and remain in that gauge for the full
evolution~\cite{Varma:2018sqd}.

We evolve Einstein's equations using a multidomain spectral
method~\cite{Kidder:1999fv, Lindblom:2005qh, Scheel:2008rj,
  Szilagyi:2009qz, Hemberger:2012jz}, with a method-of-lines
timestepper that uses a fifth-order Dormand-Prince integrator and a
proportional-integral adaptive timestepping control
system~\cite{Press:2007zz}.  The computational domain extends from
excision boundaries, located just inside the apparent horizons (AH) and
conforming to their shapes~\cite{Scheel:2008rj, Szilagyi:2009qz,
  Hemberger:2012jz, Ossokine:2013zga}, to an artificial outer boundary.
No boundary conditions are needed or imposed on the excision
boundaries.  On the outer boundary we impose constraint-preserving
boundary conditions~\cite{Lindblom:2005qh, Rinne:2006vv, Rinne:2007ui}
on most of the fields, we impose an approximate no-incoming-wave
condition on two physical degrees of freedom by freezing the
Newman-Penrose $\Psi_0$ at the boundary~\cite{Lindblom:2005qh}, and we
impose a Sommerfeld condition on the four remaining (gauge) degrees of
freedom~\cite{Lindblom:2005qh}.
The outer boundary is typically placed at a radius of order $1,000M$, but
this varies for different simulations.
For extremely long simulations, the outer boundary is
automatically placed farther outward
to avoid the center-of-mass gauge effect
reported in~\cite{Szilagyi:2015rwa}.
We also impose boundary conditions on
the incoming characteristic fields at each inter-domain
boundary using an upwind penalty method~\cite{HESTHAVEN200023,
  Bjorhus1995}.  After a common
apparent horizon forms, the simulation automatically stops,
interpolates onto a new grid with a single excision boundary inside
the new common horizon~\cite{Scheel:2008rj, Hemberger:2012jz}, and
continues evolving until after ringdown.

Spectral methods are exponentially convergent, meaning that spatial
truncation errors decrease exponentially with the number of grid
points in a particular subdomain, if the size and shape of the
subdomain remain fixed.  However, our spectral adaptive mesh
refinement (AMR) procedure~\cite{Lovelace:2010ne, Szilagyi:2014fna}
dynamically changes the size, shape, and the number of subdomains
during the simulation.  In addition, we choose the resolution not by
choosing a number of grid points, but by specifying an error tolerance
that governs when grid points should be added or subtracted in a given
subdomain ($p$-refinement), and when different subdomains should be
split and joined ($h$-refinement).  Because of this complicated AMR
procedure, we should not expect strict convergence as a function of
the AMR tolerance parameter.  To see why, consider two otherwise-identical
simulations with different AMR tolerances.  These simulations may
happen to have the same number of grid points in a particular
subdomain at a particular time, because their local truncation errors
happen to be within the appropriate thresholds. In this case they will
not exhibit strict convergence.  Similarly, strict convergence will be
lost if these two simulations happen to have entirely different
subdomain boundaries (because of $h$-refinement) at the same time.
Strict convergence can also be lost because our adaptive control
system that adjusts the size and shape of the excision
boundaries~\cite{Hemberger:2012jz} involves thresholds. One way that
this can happen is if two otherwise-identical simulations with
different AMR tolerances end up with excision boundaries in slightly
different locations. Finally, decisions by both AMR and by the control
system exhibit hysteresis. Despite these issues, most of our
simulations do exhibit convergence
with AMR tolerance, as shown in Sec.~\ref{sec:waveform_comparison}.

\subsection{Recent improvements to \software{SpEC}}
\label{sec:recent-impr-spec}

Since the time of the previous catalog paper\cite{Boyle:2019kee},
there have been numerous improvements to \software{SpEC}, with approximately 2,000 new
commits in the git history.  While some of these improvements have
been rather mundane (like upgrading to C++17 or providing support for
additional compute clusters), many of the changes have targeted
performance, robustness, and new features.

Many of the performance improvements have focused on reducing \software{SpEC}'s
memory usage by eliminating a large number of temporary variables. Given that
memory bandwidth is the limiting factor on today's HPC systems, we have also
worked on eliminating memory allocations and data copying. Other improvements
sped up computations that were prohibitively slow for certain edge cases.  An
example of this kind is the code that computes the spin of an
apparent horizon using the method of Ref.~\cite{Owen:2009sb}. This
method requires solving a generalized eigenvalue problem involving
matrices of size $\sim L^2 \times L^2$, where $L$ is the maximum $\ell$
retained in the spherical-harmonic $Y_{\ell, m}$ expansion of the
horizon surface.  Formerly, all eigenvalues and eigenvectors were
found directly, but this was prohibitively slow for simulations with
large mass ratios where $L$ of a highly distorted horizon could be 80
or more. This is now done with an iterative method that finds only the
eigenvectors with the three smallest eigenvalues, which are the ones needed to
compute the spin of the apparent horizon.

Robustness improvements were mostly driven by the need to simulate
BBHs with larger mass ratios and spins.  At the time of
Ref.~\cite{Boyle:2019kee}, most simulations had mass ratio less than
4, with higher mass ratio simulations being expensive and not infrequently
failing. Mass ratios up to about 8 are now straightforward even with
spin magnitudes of $0.8$ and precession.
Some of the improvements include better
eccentricity measurement~\cite{Habib:2024soh, Knapp:2024yww} that
handles cases where our previous algorithm failed to robustly measure the
eccentricity of the system, especially if the eccentricity was small.

A change that helps both efficiency and robustness is improvements in
the coordinate maps that connect the grid frame to the inertial
frame. In particular, several coordinate maps described in
Ref.~\cite{Hemberger:2012jz} were inverted numerically because they involved
functions like Gaussians. The numerical inversion occasionally
failed for edge cases (e.g., unlucky machine roundoff
caused the wrong root to be found), and more of these failures happened
for high mass ratios,
eccentric systems, and hyperbolic encounters, when the map parameters became
more extreme.  Although these failures were rare, we often run hundreds
of simulations at once, so diagnosing and fixing each new edge case took
considerable human time.  The solution was to replace these maps
with new maps that use piecewise linear functions with kinks at subdomain
boundaries so that the maps are quicker to compute, more accurate, and can be
inverted analytically. Restricting the kinks to subdomain boundaries retains
spectral convergence, since the map inside each subdomain remains
smooth. Additionally, several bugs and inaccuracies in the time-dependent
coordinate maps were fixed. These issues do not appear to
impact the waveforms.

A failure mode that rarely but regularly appeared either at the beginning of the
simulation (for high spins) or at the beginning of ringdown was caused
by a mismatch between the shape of the excision boundaries and the
shape of the horizons. The origin of this problem
is the use of different grids before
and after the transition, either from initial data to inspiral or inspiral to
ringdown.  This mismatch sometimes prevented the control
system from locking and the simulation failed after a few timesteps.
Now, if such a failure occurs, the horizon shape is measured during
the first few timesteps and then the simulation is automatically restarted
with new excision boundaries that better match the horizon shapes. This process
is repeated several times if necessary, typically up to three.

There have also been many improvements to the initial data solver since
Ref.~\cite{Ossokine:2015yla}, primarily to make the solver more robust for very
large initial distances between the two black holes.  Among the changes are (i)
a radial split of the inner spherical shells surrounding either excision
boundary, if the radial coverage is too large; (ii) streamlining the criteria
used to decide when to increase the AMR resolution and how to measure the error
in root-finding; (iii) fixes to bugs that are triggered by initial data sets
with particularly large separations or BH velocities; (iv) changes to the
preconditioning that lead to more robust and more efficient solution of the
discretized linear system; (v) the possibility to specify ADM energy and angular
momentum of the initial-data set instead of orbital frequency and radial
velocity; and (vi) various code-cleanup and refactoring to improve
maintainability. In addition, because \software{SpEC} was originally written and
tested with binaries on quasicircular orbits in mind, running simulations of
highly eccentric and hyperbolic-like orbits required many adjustments, which we
outline in Sec.~\ref{sec:eccentric_hyperbolic}.

We have also implemented different initial data that provide improvements for
reducing junk radiation and improving performance of high-spin simulation.  For
newer simulations we typically use SHK~\cite{Varma:2018sqd}
initial data for spins below 0.8 and SphSKS~\cite{Chen:2021rtb}
for larger spins. Which type of initial data was used is available in the
metadata field \texttt{initial\_data\_type} (see
Appendix~\ref{sec:metadata_fields} for metadata details). We also typically use
a negative expansion boundary condition in the initial data.
This allows us to
place the excision boundary inside the apparent horizon in the initial data
and eliminates the need to extrapolate to the horizon interior when starting the
evolution~\cite{Varma:2018sqd}.

New features include a new method for measuring
spin directions based on Ref.~\cite{Owen:2017yaj},
and new methods of constructing free fields for initial
data~\cite{Varma:2018evz,Chen:2021rtb}. Additionally, we have tuned AMR
tolerances to produce a more uniform error across the grid and throughout the
simulation. In particular, we have increased the resolution in the wavezone
spherical shells to ensure that they are not the dominant source of error. As
part of this, we improved the way we handle spherical harmonic
filtering. \software{SpEC} uses a scalar spherical harmonic decomposition, which
causes aliasing errors when involving higher rank tensors. A tedious but
straightforward derivation shows that this aliasing can be eliminated by
transforming to a tensor spherical harmonic basis, zeroing the top four $\ell$
modes, and then transforming back to the scalar spherical harmonic basis~\cite{Scheel:2006gg}. We
precompute and cache the transformation rules on disk, then load them for the
current $L$ in each spherical shell. This storage algorithm
has seen several improvements, including copying the cache on disk
to node-local SSDs to reduce load time.

Improvements to the feedback control systems~\cite{Scheel:2008rj,
  Hemberger:2012jz, Scheel:2014ina} were required for robustly
simulating higher mass ratio configurations, mostly requiring
tightening control system tolerances to avoid drift from the optimal
solution. A major overhaul of the rotation control system was also
done, so that it now has the same interface as the other control
systems.  This eliminated many special cases in the control system
code, reducing the complexity of that code considerably.

Extracting all the Weyl components and extrapolating them to infinity is
extremely inaccurate unless junk radiation is very small.  We can
significantly reduce the junk radiation by allowing the
initial junk radiation pulse to evolve to the
outer spherical shells, and then dropping the shells containing the
junk radiation~\cite{Iozzo:2020jcu}. This causes a discontinuous
change in the outer boundary, but drastically reduces junk radiation
reflections off the outer boundary. While effective, this procedure
is not done by default in our simulations.

Finally, two key new features are
a way to reduce the effect of junk radiation on
errors, as explained in detail in Sec.~\ref{sec:pbandj}, and a new algorithm
for determining the end of the junk phase (the relaxation time), as described in
Sec.~\ref{sec:reference-time-algorithm}.

\subsection{Perform branching after junk}
\label{sec:pbandj}

Assessing the convergence of numerical relativity simulations is complicated by the
presence of initial transients, usually referred to as junk
radiation. We intentionally configure AMR so that it does not
resolve the initial junk transients. This is for
two reasons: First, the junk radiation is not
astrophysically interesting, so the beginning of the simulation that
contains significant junk is eventually discarded.  Second, resolving
the junk is computationally expensive because the transients have
small wavelengths and large frequencies.

Because the junk is not resolved, two otherwise-identical simulations
starting at $t=0$ with different AMR tolerances have different,
nonconvergent junk transients. As a result, they have slightly
different masses, spins, and orbital parameters even at
$t=t_\mathrm{junk}$, where $t_\mathrm{junk}$ is some time after which
the primary junk transients are deemed to have sufficiently decayed
away, though residual reflections may remain (see
Sec.~\ref{sec:reference-time-algorithm}).  In other words, the
unresolved junk transients leave behind an effectively random small
change in the physical observables. Because of this,
otherwise-identical simulations starting at $t=0$ with different AMR
tolerances can fail to converge even at times $t>t_\mathrm{junk}$.

Our solution to this problem is to modify the way in which we specify
simulations with multiple resolutions.
We call this method ``perform branching after junk''
(PBandJ). Instead of starting multiple simulations with
different AMR tolerances at $t=0$, we start a single simulation with
one AMR tolerance at $t=0$.  We then choose some time
$t_\mathrm{PBandJ} > t_\mathrm{junk}$, which is independent of the
reference time chosen by the algorithm in
Sec.~\ref{sec:reference-time-algorithm}.
When that simulation gets to
$t=t_\mathrm{PBandJ}$ we fork several copies of that simulation, each
copy having a different AMR tolerance for $t>t_\mathrm{PBandJ}$.
Effectively, a snapshot of the first simulation at $t=t_\mathrm{PBandJ}$
serves as initial data for the copies.  This procedure provides better
convergence results and error estimates, by simply ignoring the
initial part of the simulation that is astrophysically uninteresting
and that we will discard anyway for waveform analysis.
Most simulations in the catalog carried out after year 2020 use this
procedure, and $t_\mathrm{PBandJ}$ is typically chosen to be about 2.5
orbits after $t=0$. Typically, the first simulation that evolves
from $t=0$ to $t=t_\mathrm{PBandJ}$ is the highest-resolution simulation,
and the forked copies are at lower resolution.  However,
occasionally we decide that our resolution is insufficient after the
first simulation has finished, and in this case
we fork an additional copy at higher resolution starting at
$t=t_\mathrm{PBandJ}$.

\subsection{Adjustments for highly eccentric and hyperbolic-like orbits}
\label{sec:eccentric_hyperbolic}

To evolve highly eccentric ($e_{\mathrm{ref}}\gtrsim 0.5$) BBHs and hyperbolic-like encounters in
\software{SpEC}, several adjustments are made to resolve the fast
dynamics characteristic of these systems.
Because eccentric binaries emit pulses of GWs, the algorithm to choose
the output
frequency of extracted GWs was changed to be proportional to the highest
instantaneous orbital frequency achieved during the entire past of the
simulation.
For hyperbolic events, it is not possible to meaningfully define
an orbital frequency,
so instead waveforms are exported in timesteps of $0.5M$, which we
find to be sufficient to capture the characteristic burst of radiation
emitted during the periastron passage.

Junk radiation requires special care for these types of simulations, since we usually do not have extra early orbits that we can discard.
Instead, scattering and highly eccentric encounters start at very large initial distance (typically, $D_0=250M$) to make sure that junk radiation has sufficiently decayed before the first periastron passage.
Junk radiation is also what necessitated some of the changes to the initial data
code detailed above.
Previous versions of \software{SpEC} suppressed AMR in the
outer spherical shell grids until
the pulse of junk radiation passed through the outer boundary.
However, because of the fast dynamics of hyperbolic or extremely-eccentric
cases, inner shells need refinement
already fairly early on,
usually before the pulse crosses the outer boundary.
To allow inner shells to refine as soon as possible, AMR
is no longer suppressed globally,
but individual spectral elements are allowed to refine as soon
as the junk pulse passes through them.

Another adjustment for these systems
is the choice of shorter time scales to trigger AMR, which is necessary
for AMR to adapt quickly enough to the rapid change in the radial separation.
Furthermore, significant refinement is needed during the close periastron
passage. 
We have therefore added the ability for AMR to split (i.\,e.\ $h$-refine) 
hollow cylindrical elements in the angular direction.

The gauge source functions in \software{SpEC} start from an initial
gauge that is slowly rolled off towards the damped harmonic gauge.
If the initial separation between the black holes is very small,
we shorten this rolloff time to make sure the system is in purely
damped harmonic gauge well before the first periastron passage.
Likewise, the time for PBandJ
(see Sec.~\ref{sec:pbandj}) is chosen such that it occurs
before the first periastron passage. Typically, junk radiation
is still present in the outer shells at the PBandJ time, but the most severe
impact of junk radiation is mitigated.
More details on the adjustments for highly eccentric and hyperbolic-like orbits
will be presented in~\cite{Long:HyperbolicSimsInPrep}.

\subsection{Eccentricity and mean anomaly}
\label{sec:eccentricity}

Each simulation in the catalog contains metadata fields
for eccentricity and mean anomaly (see Appendix~\ref{sec:metadata_fields}
for a list of all metadata fields.).  Because there are multiple definitions
of eccentricity in general relativity, and because both eccentricity
and mean anomaly vary as the simulation proceeds, we emphasize that
users who care about particular definitions and precise values of
these quantities should compute eccentricity and mean anomaly
themselves in a consistent manner, using either waveforms or
trajectories according to whatever method they choose.  In particular,
when comparing the eccentricities and mean anomalies of a series of
simulations, those comparisons are most physically meaningful when
the eccentricities and mean anomalies are computed at some time
relative to merger. In contrast, here we compute eccentricities and
mean anomalies at some time early in the inspiral after junk, since
the number of orbits in our simulations varies widely from case to
case.

Nevertheless, we describe here how we compute eccentricities and mean
anomalies.  We use three methods.  The first is the method described
in Ref.~\cite{Habib:2024soh}, where a post-Newtonian-based functional
form (Eq.~(10) of~\cite{Habib:2024soh})
is fit to the time derivative of the angular frequency of the
orbital trajectory.  This functional form is based on an expansion in
small $e$, and includes spin-spin corrections and radiation reaction.

The second method is the one described in Ref.~\cite{Knapp:2024yww},
which is similar to the first method, except that the functional form
(Eq.~(9) of~\cite{Knapp:2024yww}) allows for general $e$ that can be
large.  For general $e$, it is usually necessary to invert the Kepler
equation to solve for the eccentric anomaly $u(t)$ as a function of
time $t$. Here the Kepler equation at 1PN order reads
\begin{align}
  t - t_{\mathrm{ref}}&= a^{3/2} M^{-1/2}
  \left(1 + \frac{9 - \eta}{2a M^{-1}} \right) \left[u(t)-e \left(1 + \frac{3 \eta - 8}{2a M^{-1}}\right)\sin u(t)-\ell\right],
  \label{eq:Kepler}
\end{align}
where $e$ is the (Newtonian) eccentricity, $\ell$ is the mean anomaly at time
$t=t_{\mathrm{ref}}$,
$a$ is the (Newtonian) semimajor axis, $M$ is the total mass $m_1+m_2$,
and $\eta$ is the symmetric mass
ratio $m_1 m_2/(m_1+m_2)^2$. Here we use geometric units $G=c=1$.
However, in Ref.~\cite{Knapp:2024yww} the inverse of the Kepler equation
is approximated as the closed-form expression
\begin{align}
  u(t) &= a^{-3/2} M^{1/2}
  \left(1-\frac{9-\eta}{2a M^{-1}}\right) (t - t_{\mathrm{ref}}) + \ell,
  \label{eq:KeplerApprox}
\end{align}
which is a reasonable approximation for $e\lesssim 0.5$, but not for
larger eccentricities.

The third method works best for large eccentricities, when
eccentricity dominates the trajectory.  This method is the same
as the second method, except the functional form that is fit to the
trajectory omits spin-spin terms and radiation-reaction terms, and the
Kepler equation is inverted numerically.  In other words, the functional
form is Eq.~(9) of~\cite{Knapp:2024yww} with $C_1=C_2=C_3=C_4=0$, with $u(t)$
determined numerically from Eq.~(\ref{eq:Kepler}).

We use the first method for all cases with $e < 0.01$. For very small
eccentricities the mean anomaly is degenerate and has no physical
meaning. For larger eccentricities $e > 0.01$, we try both the second
and third method, and choose the result with the smallest value of
the $L^2$ norm of the fit residual divided by the number of fit parameters.

Note that there are some simulations with empty eccentricity or mean
anomaly metadata fields, and other simulations where the eccentricity
metadata field is a string (where the string describes a reason for lack of a
numerical eccentricity such as ``simulation too short'').  These
simulations are either head-on or nearly-head-on collisions,
hyperbolic encounters where the objects are unbound and escape to
infinity, or simulations with too few orbits to reliably measure an
eccentricity.

\subsection{Junk radiation, relaxation time, and reference time.}
\label{sec:reference-time-algorithm}

Although users are free to use their own methods to remove
initial ``junk radiation'' transients from simulations, we provide
several metadata fields to assist with this.  The metadata provides
a suggested \texttt{relaxation\_time} in units of total mass $M$.
This is our estimate of the amount of time that should be removed
from the beginning of the time series so that junk remains
acceptably small.  Note that \texttt{relaxation\_time} is
simulation-dependent and resolution-dependent. Users with
applications that are particularly sensitive to junk are encouraged
to measure the junk themselves in a manner of their choosing
and truncate the time series
appropriately for their use case.

We recently presented a new algorithm for determining
\texttt{relaxation\_time}~\cite{Pretto:2024dvx}. This method, which we
call HHT for Hilbert-Huang Transform~\cite{Huang:1998}, involves
constructing and analyzing the empirical mode
decomposition~\cite{Huang:1998} of a signal and determining when
high-frequency content has decayed to a desired tolerance. In our
case, the signal we use is the irreducible horizon mass as a function
of time, because it exhibits junk effects and settles to a constant at
late times. This HHT method is used for most of our simulations.
However, the new algorithm sometimes fails
for some shorter and older simulations.  For example, sometimes the
simulation is so short that junk does not fully decay before the simulation
ends; this is especially problematic for head-on and scattering
simulations.  When the HHT method fails,
we revert back to the old RMS (root-mean-squared) algorithm to determine the
\texttt{relaxation\_time} (see
\cite{Pretto:2024dvx} for a description of failure modes and a
description of the RMS algorithm).

The metadata for each simulation provides another field called
\texttt{reference\_time}.  This is the time at which we measure
quantities that parameterize the simulation, such as the individual
black-hole masses and spins, and orbital parameters such as
coordinate separations and orbital eccentricity.  The values of
these quantities at \texttt{reference\_time} are different from
their values at $t=0$ both because of junk radiation and because
some of these quantities (e.g., spin directions and eccentricity)
have non-negligible time-dependence even early in the simulation.
Typically we choose \texttt{reference\_time} equal to
\texttt{relaxation\_time}.  However, for some simulations (typically
very short ones such as nearly head-on collisions) we choose
\texttt{reference\_time} by hand.  The catalog metadata for each
simulation now includes information as to which algorithm was used
to determine the relaxation and reference times, in the metadata
field \texttt{t\_relaxed\_algorithm}.

\subsection{Memory correction}
\label{sec:memory_correction}

An important prediction of Einstein's theory of relativity is that
whenever a system emits gravitational radiation, that radiation will
permanently change the spacetime that it propagates through. This
permanent change is called the \emph{gravitational wave memory} and
predominantly manifests in a gravitational wave as a net change in the
strain between early and late times. Waveforms that are extracted at
future null infinity from numerical relativity simulations using
extrapolation, however, fail to naturally
capture memory~\cite{Favata:2008yd}.
Fortunately, Ref.~\cite{Mitman:2020bjf} showed that these waveforms
can be corrected to include memory by computing and adding certain
contributions to the strain that seem to be missing by using the
BMS balance laws. With this
correction, extrapolated waveforms respect the Bianchi identities to a
much higher degree and agree much more with waveforms extracted using
Cauchy-characteristic evolution
(CCE)~\cite{1996PhRvD..54.6153B, Bishop:2016lgv, 2016CQGra..33v5007H,
  2020PhRvD.102b4004B, 2020PhRvD.102d4052M, Moxon:2021gbv} that naturally
contain the memory. Consequently, we have updated the waveforms in our
catalog to include memory, using the technique outlined in
Ref.~\cite{Mitman:2020bjf}.

The procedure consists of calculating the null memory contribution to the
strain via
\begin{align}
	\label{eq:memory_correction}
	h^{\mathrm{memory}}=\frac{1}{2}\bar{\eth}^{2}\mathfrak{D}^{-1}\left[\frac{1}{4}\int_{-\infty}^{u}|\dot{h}|^{2}du\right],
\end{align}
where
\begin{align}
	\eth f(\theta,\phi)\equiv-(\sin\theta)^{+s}\left(\partial_{\theta}+i\csc\theta\partial_{\phi}\right)\left[\left(\sin\theta\right)^{-s}f(\theta,\phi)\right]
\end{align}
is the Geroch-Held-Penrose spin-weight operator (here represented when
acting on a spin-weight $s$ function $f$ in spherical
coordinates)~\cite{Geroch:1973am} and
\begin{align}
	\mathfrak{D}\equiv\frac{1}{8}\left(\bar{\eth}\eth\right)\left(\bar{\eth}\eth+2\right).
\end{align}

For each of our simulations, we correct each waveform by computing
$h^{\mathrm{memory}}$ via Eq.~\eqref{eq:memory_correction} and adding
it to the strain. Note that we take the lower limit of integration to
be the relaxation time
of the simulation. We also correct $\Psi_{4}$
by adding $-\frac{1}{2}\ddot{h}^{\mathrm{memory}}$.

\section{Data archive, versioning for reproducibility, and user interface}
\label{sec:data_management}

As the number of simulations in the SXS catalog continues to grow, the
management of the data becomes increasingly complex and challenging.
To support new analyses and to ensure reproducibility of existing
results, the data must be made available to the scientific community
in ways that are easy to search and access, while scaling to
accommodate the growing number of simulations in the catalog, yet
remaining accessible over the long term.  To balance these sometimes
conflicting objectives, we have developed a user interface that
provides consistency while transparently handling a variety of data
formats and storage locations.

\subsection{Archiving and versioning data}
\label{sec:archiving}

We assign a unique ``SXS ID'' to each simulation in the catalog, of
the form \texttt{SXS:BBH:1234}.  The numerical portion is---as
yet---always a four-digit number, zero-padded if necessary.  The
numbers are not always consecutive, and may not correspond to when the
simulations were performed.  Each simulation is deposited as a
separate record in a long-term open-access digital
repository---originally Zenodo~\cite{Zenodo}, and more recently
CaltechDATA~\cite{CaltechDATA}.  Corresponding to the ID, each
simulation is also given a digital object identifier (DOI) that is
simply the SXS ID combined with the SXS Collaboration's DOI prefix,
\texttt{10.26138}.  Thus, for example, SXS:BBH:1234 will always be
accessible at \url{https://doi.org/10.26138/SXS:BBH:1234}, which
redirects to whichever digital repository holds that simulation.

However, the data for a simulation may be updated over time, as
improvements are made to the post-processing methods, as conventions
change, or as bugs are fixed.  We have made such updates a number of times in the past.  One of the more significant
changes was an overall sign change to the definition of the strain,
introduced just before the 2019 catalog~\cite{Boyle:2019kee} and
applied retroactively to simulations run before 2019.  That update to
the catalog also introduced center-of-mass
corrections~\cite{Woodford:2019tlo}.  The update of the catalog
associated with the current paper introduces memory corrections to the
waveforms, as discussed in Sec.~\ref{sec:memory_correction}.  We have
applied memory corrections and center-of-mass corrections to all
simulations, not only newer ones. There have been numerous less
significant changes, such as small corrections to metadata.

Any time a file changes in any way, it could potentially alter the
results of some analysis.  To ensure true reproducibility, we now use
version numbers for each simulation, which are incremented whenever
any file is changed.\footnote{The number of versions varies between
simulations, ranging from 8 for some older simulations to just 1 for
the newest simulations.}  As of this release of the catalog, all
simulations have version number 3.0; prior releases have also been
specified with lower version numbers.  These numbers can be appended
to the SXS ID to specify the version.  In particular, each version of
each simulation is also given a DOI, such as
\texttt{10.26138/SXS:BBH:1234v3.0}.  While the unversioned DOI will
always be updated to point to the most recent version, the versioned
DOIs will always point to the specific data associated with that
version.  For reproducibility, it is best to specify the version
number whenever possible.

Referring individually to versioned SXS IDs can be cumbersome.
Therefore, we also provide an overall version number for the
collection of the most recent versions of all simulations at any time.
As of the release of this paper, the catalog version is 3.0.0, and is 
archived on Zenodo~\cite{SXSCatalogData_3.0.0}.  This version
number will be incremented whenever any simulation is updated.  By
referring to the catalog version, the version of any particular
simulation---unless otherwise stated---is implied by what is stored in
that version of the catalog.  We recommend that users specify the
catalog version when discussing analyses that use simulations from a
single version the catalog, or specific simulation versions otherwise.
These versions should also be cited, for which we provide a convenient
function \texttt{sxs.cite} which can provide the citation for a
particular catalog version, as with Ref.~\cite{SXSCatalogData_3.0.0},
or citations for specific simulations and the papers that introduced
them.
We also provide a user interface that directly uses these SXS IDs,
with or without versions, and the versioned catalog.

\subsection{User interface}
\label{sec:user_interface}

In constructing the user interface, we begin by considering the perspective of a new user approaching the
catalog.  First, the user needs to know what data are available, and
be able to easily sort through the simulations to find those that are
relevant to a particular analysis.  While this paper provides a
high-level aggregate survey, the user will need detailed information
about each simulation individually, and means of selecting them both
interactively and algorithmically.  Having selected the simulations to
analyze, the user will then need to know where and how to obtain the
data, as well as how to load and use the data.  Ideally, the procedure
should be as simple and uniform as possible, remaining constant over
time so that analyses can be reliably reproduced---or expanded and 
reused---in the future.

To meet these needs we provide the \software{sxs} Python
package,\footnote{The \software{sxs} package can be installed either
using \software{pip}/\software{uv} or \software{conda}/\software{mamba}/\software{pixi}.  The 
source code and documentation are available at
\url{https://github.com/sxs-collaboration/sxs}.  Each release is
archived at \url{https://doi.org/10.5281/zenodo.13891077}.  }  %
which can automatically obtain and load the metadata for the entire
catalog, as well as data for each simulation individually.  The
metadata can be loaded into a single table, which can be filtered and
sorted using standard pandas queries to find simulations of interest.
We provide an interactive interface to the metadata at
\url{https://data.black-holes.org}.\footnote{This interface is based
on a \texttt{marimo} notebook~\cite{Marimo:2023}, which uses
\texttt{Pyodide}~\cite{Pyodide:2025} to run a simplified version of
the \software{sxs} package in a Python session directly in the browser.}
The metadata can be loaded, for example, with
\texttt{sxs.load("simulations", tag="3.0.0")}, where the optional
\texttt{tag} argument specifies the version of the catalog to load.
Because the metadata are heterogeneous and not necessarily ideal for
searching or sorting, it is also possible to load a
\texttt{"dataframe"} with homogeneous data types for each field.  In
either case, the chosen version of the catalog will then provide the
default version of simulations throughout the Python session.

Any particular simulation can be loaded by SXS ID---for example as
\texttt{sxs.load("SXS:BBH:1234")} or \texttt{sxs.load("SXS:BBH:1234v3.0")}.  Data available for each simulation
include horizon information as well as strain and $\Psi_4$ waveforms,
and can be ``lazily'' obtained and loaded only when needed, simply by
accessing attributes of the simulation object.  This frees the user
from having to download and manage the data manually---or even know
where or in what format the data are stored.  By default, the highest
available resolution is used.  In the case of extrapolated waveforms,
a default extrapolation order of $N=2$ is chosen.  However, different
choices can easily be specified by the user. The package also caches
data locally, speeding up subsequent accesses, though this
behavior can also be overridden by the user.  See
Listing~\ref{lst:sxs_package_example} for an example of using the
\software{sxs} package.

\begin{lstlisting}[
  float=tp,
  floatplacement=tbp,
  basicstyle=\ttfamily\small,
  language=Python,
  keywordstyle=\color{ForestGreen},
  stringstyle=\color{Maroon},
  commentstyle=\color{gray},
  showstringspaces=false,
  columns=fullflexible,
  caption={An example of using the \software{sxs} Python package to load
  data.  Note that the user does not have to download any data; the
  package will automatically manage downloading and caching data.},
  label=lst:sxs_package_example]
  import sxs
  import matplotlib.pyplot as plt

  # Load a pandas.DataFrame of metadata for all simulations
  df = sxs.load("dataframe", tag="3.0.0")

  # Find SXS ID of BBH simulation with highest effective spin
  highest_chi_eff = df.BBH["reference_chi_eff"].idxmax()

  # Load that simulation
  sim = sxs.load(highest_chi_eff)

  # Print a summary of the simulation's parameters
  print(sim)

  # Print all metadata of the simulation
  print(sim.metadata)

  # Obtain and load the strain data
  h = sim.h

  # Plot the real parts of the (2,2) and (2,0) modes
  plt.plot(h.t, h.real[:, h.index(2,2)], label=r"(2,2)")
  plt.plot(h.t, h.real[:, h.index(2,0)], label=r"(2,0)")
  plt.xlabel("Time ($M$)")
  plt.ylabel(r"$\Re\{h^{\ell, m}\}$")
  plt.title(sim.sxs_id)
  plt.legend()
\end{lstlisting}

Partly because of the number of versions of each simulation, and
partly just because of the increasing number of simulations, the sheer
size of the catalog has become difficult to sustain.  The total size
of all simulations prior to this data release was \oldtotalsize.  A
user who wanted just one waveform from the highest resolution of each
simulation needed to download and/or store over \oldwaveformsize of
data for all \oldnumberofsimulations simulations in the previous
version of the catalog.

We use several techniques to deal with this problem. First, we are
limiting the types of data we publish to Zenodo or
CaltechDATA---though we retain all data locally.
Previously, we included finite-radius waveforms from the
simulations, alongside the extrapolated waveforms.  We have always
advised against using finite-radius waveforms, as we expect the extrapolated
waveforms to be more physically relevant.  Moreover, there are roughly
7 times more finite-radius waveforms than extrapolated, making the
catalog far larger than it needs to be.  Therefore, we no longer publish
finite-radius waveforms as part of the catalog.  Second, we have introduced a
new waveform format that compresses each waveform by an average factor of 7
compared to the old (compressed but wasteful) format.  As described in
Appendix~\ref{sec:waveform_format}, the new format applies a number of
complicated non-standard transformations to the data.

Together, these changes have reduced the size of this most recent
version of the catalog to just \newtotalsize in total.  A user wanting one
waveform from the highest resolution of each simulation will need to
download and/or store just \newwaveformsize for all
\numberofsimulations simulations---\bestwaveformratio times less space per
simulation.  This dramatic reduction in size
improves the user experience substantially, from downloading the data
to storing it.  However, because of the admittedly cryptic waveform
format, it is unrealistic to expect the user to load the data directly
from the supplied files.  Instead, the \software{sxs} package is
designed to insulate the user from this challenge and other
inconveniences such as changes to file names and locations, or even 
future changes to formats.

\section{Conclusion}\label{sec:conclusion}
In this paper, we have presented an update to the Simulating eXtreme Spacetimes
(SXS) BBH catalog. Using the highly efficient spectral methods
implemented in the Spectral Einstein Code (\software{SpEC}), we have increased the total
number of configurations from \oldnumberofsimulations to \numberofsimulations. The catalog now
densely covers the parameter space with precessing simulations up to mass ratio $q=8$ with dimensionless
spins up to $|\chi|\le0.8$ with near-zero eccentricity. The catalog includes some simulations at
higher mass ratios with moderate spin and more than 250 eccentric
simulations. We have also deprecated and rerun older
simulations and ones with anomalously large errors in the waveform. The median waveform difference
between resolutions over all \numberofsimulations simulations is
$4\times10^{-4}$, with a median of
\mediannumberoforbits orbits, while the longest simulation is
\largestnumberoforbits orbits. All the
waveforms in the catalog are center-of-mass and gravitational-wave-memory
corrected~\cite{Woodford:2019tlo, Mitman:2020bjf}. We
provide a python package, \software{sxs}~\cite{sxs}, to simplify accessing the
catalog. Because of a new compression algorithm and deprecation of lower-quality
simulations, the \software{sxs} package is by far the best method for users to
access the catalog. We estimate the total CPU cost of
all the simulations in the catalog to be
only \simscharged core-hours.
Using spectral methods for long, precessing BBH
inspiral-merger-ringdown simulations is over 1,000 times more efficient than
using finite-difference methods for a few orbits of non-spinning BBHs at
comparable accuracies; see, e.g.,~\cite{Rashti:2024yoc}. To date GPU-based 
finite-difference codes have yet to prove
competitive with CPU-based codes using spectral methods.
The full catalog is publicly available at \url{https://data.black-holes.org}\,.

We assess the readiness of our catalog for use in current and next-generation
detectors in Sec.~\ref{sec:waveform_comparison}. We find that for simulations long
enough to span the entire signal in the detector band, most simulations are
accurate enough for current detectors, but significant improvements need to be
made, both in terms of accuracy and length, for next-generation ground-based and
space-based observatories. Significant improvements will also need to be made
for next-generation detectors in hybridizing
post-Newtonian and numerical relativity waveforms~\cite{Santamaria:2010yb,
  MacDonald:2011ne, Boyle:2011dy, MacDonald:2012mp, Varma:2018mmi,
  Sadiq:2020hti, Mitman:2021xkq, Mitman:2022kwt, Sun:2024kmv}
in order to have waveform models that span
the entire signal that will be in band.

In the future, we will expand the catalog towards higher mass ratios and focus
on filling out the eccentric parameter space. More challenging will be
increasing the accuracy to meet the requirements of next-generation detectors,
especially since the required length of numerical-relativity waveforms depends
on how late one can hybridize them with post-Newtonian waveforms. A major
challenge for increasing accuracy is significantly reducing the spurious
``junk'' gravitational radiation generated by imperfect initial data (see,
e.g., Ref.~\cite{Varma:2018sqd} for some progress in this direction). Finally, in the
near term we will release a catalog of all of the simulations presented here,
but using Cauchy-Characteristic Evolution~\cite{1996PhRvD..54.6153B,
  2016LRR....19....2B, 2016CQGra..33v5007H, 2020PhRvD.102b4004B,
  2020PhRvD.102d4052M} to extract the GWs. This has several advantages, such as naturally producing the correct
gravitational wave memory and independently producing all five complex Weyl
components, the news, and the strain at future null infinity without any need to
differentiate or integrate quantities.

\ack
We thank Zachariah Etienne and Sebastian Khan for pointing out problems
with some of the now-deprecated waveforms.
We thank Tom Morrell at the Caltech Library for his advice on hosting
the data archive with CaltechDATA, and we thank the Caltech Library
for providing this excellent service.
This material is based upon work supported by the National Science Foundation
under Grants No.~PHY-2309211, No.~PHY-2309231, and No.~OAC-2209656 at Caltech; %
by No.~PHY-2407742, No.~PHY-2207342, and No.~OAC-2209655 at Cornell; %
by CAREER Award No.~PHY-2047382 at the University of Mississippi; %
by No.~PHY-2309301, DMS-2309609, PHY-2110496, and AST-2407454 at the University of Massachusetts, Dartmouth; %
and by No.~PHY-2208014 and AST-2219109 at Cal State Fullerton.
V.V.~and S.F.~acknowledge support from University of Massachusetts, Dartmouth's Marine and Undersea Technology (MUST) Research Program funded by the Office of
Naval Research (ONR) under Grant No. N00014-23-1-2141.
Any
opinions, findings, and conclusions or recommendations expressed in this
material are those of the author(s) and do not necessarily reflect the views of
the National Science Foundation. This work was supported by the Sherman
Fairchild Foundation at Caltech and Cornell; %
by a Sloan Foundation Research Fellowship at the University of Mississippi; %
and by  the Dan Black Family Trust, and Nicholas and Lee Begovich at Cal State
Fullerton.
Support for this work was provided by NASA through the NASA Hubble Fellowship
grant number HST-HF2-51562.001-A awarded by the Space Telescope Science
Institute, which is operated by the Association of Universities for Research in
Astronomy, Incorporated, under NASA contract NAS5-26555.
H.\,R.\,R.\@ acknowledges financial support provided
under the European Union's H2020 ERC Advanced Grant ``Black holes:
gravitational engines of discovery'' grant agreement no.~Gravitas–101052587.
Views and opinions expressed are however those of
the authors only and do not necessarily reflect those of the European
Union or the European Research Council.  Neither the European Union
nor the granting authority can be held responsible for them.
A. R.-B. is supported by the Veni research programme which is financed by the Dutch Research Council (NWO) under the grant VI.Veni.222.396; 
acknowledges support from the Spanish Agencia Estatal de Investigación grant PID2022-138626NB-I00 funded by MICIU/AEI/10.13039/501100011033 and the ERDF/EU; 
is supported by the Spanish Ministerio de Ciencia, Innovación y Universidades (Beatriz Galindo, BG23/00056) and co-financed by UIB.
K. Z. C. acknowledges support by the Hungarian Scientific Research Fund NKFIH Grant No. K-142423.

Computations were performed on the Wheeler cluster at Caltech, which is
supported by the Sherman Fairchild Foundation and by Caltech. The computations
presented here were conducted in the Resnick High Performance Computing Center,
a facility supported by Resnick Sustainability Institute at the California
Institute of Technology.  This work used Anvil at Purdue
University~\cite{Anvil}, Expanse at San Diego Supercomputer
Center~\cite{Expanse}, Stampede 2 at Texas Advanced Computing Center, through
allocation PHY990002 from the Advanced Cyberinfrastructure Coordination
Ecosystem: Services \& Support (ACCESS) program, which is supported by
U.S. National Science Foundation grants \#2138259, \#2138286, \#2138307,
\#2137603, and \#2138296~\cite{NsfAccess}. The authors acknowledge the Texas
Advanced Computing Center (TACC) at The University of Texas at Austin for
providing computational resources that have contributed to the research results
reported within this paper. URL:~\url{http://www.tacc.utexas.edu}. This work
used the Extreme Science and Engineering Discovery Environment (XSEDE), which is
supported by National Science Foundation grant number ACI-1548562. Specifically,
it used the Bridges-2 system, which is supported by NSF award number
ACI-1928147, at the Pittsburgh Supercomputing Center (PSC), and the Comet
system, which is supported by NSF award number 1341698, at the San Diego
Supercomputer Center~\cite{NsfXsede}. This work is also part of the
NCSA Blue Waters PRAC allocation support by the National Science Foundation
(award number OCI-0725070)~\cite{Bode2013, Kramer2015}.
Computations were performed on the high-performance 
computer system Minerva at the Max Planck Institute for Gravitational 
Physics in Potsdam, as well as on the Raven and Urania clusters at the 
Max Planck Computing and Data Facility. Computations were performed on 
the Ocean high-performance computing cluster at Cal State Fullerton. 
Computations were performed using the Dutch national e-infrastructure
 with the support of the SURF Cooperative using grant no.  EINF-7366 and NWO-2024.002.

\appendix

\section{Metadata fields}
\label{sec:metadata_fields}

Using the $\software{sxs}$ package, metadata fields can be obtained as a
python dictionary via $\texttt{sim.metadata}$ for an individual
simulation $\texttt{sim}$, or as entries in a pandas DataFrame for all
simulations via $\texttt{sxs.load(``dataframe'')}$, as illustrated in
Listing~\ref{lst:sxs_package_example}. The ``introduced'' version marks the
first \texttt{metadata\_format\_revision} in which a metadata field is
available.

\newcommand{\metadataGroup}[2]{%
\subsection*{Metadata field group: #2}
}

\newtheoremstyle{mdFieldStyle}%
{3pt}%
{3pt}%
{}%
{}%
{\bfseries}%
{}%
{ }%
{\texttt{\thmnote{#3}}}%

\theoremstyle{mdFieldStyle}

\newtheorem*{metadataFieldInner}{Field}

\newenvironment{metadataField}[5]{%
\metadataFieldInner[#1] (type: \texttt{#2}; introduced: v#4\ifstrempty{#5}{}{; deprecated: v#5}) \newline
\hangindent=1em
}{%
\endmetadataFieldInner
}
\metadataGroup{id}{Identification}
\begin{metadataField}{simulation\_name}{str}{id}{0}{}
A non-unique SXS-assigned identifier chosen before the simulation has been run. Useful only for SXS members building and debugging the catalog.
\end{metadataField}
\begin{metadataField}{alternative\_names}{list[str]}{id}{0}{}
Comma-separated array of alternative names, longer, more descriptive, and/or indicating the specific series of simulations this configuration belongs to.  One of these alternative names is the `SXS:BBH:dddd' id-number, which is guaranteed to be unique.
\end{metadataField}
\begin{metadataField}{keywords}{list[str]}{id}{0}{}
List of free-form keywords.  Presence of the keyword `deprecated' means that this simulation has been deprecated.
\end{metadataField}
\begin{metadataField}{job\_archiver\_email}{str}{id}{2}{}
Email of person who archived this simulation into the catalog; usually the person who ran the simulation. Useful only for SXS members building and debugging the catalog.
\end{metadataField}

\metadataGroup{refs}{References}
\begin{metadataField}{citation\_dois}{list[str]}{refs}{2}{}
DOIs to cite when using this simulation.
\end{metadataField}

\metadataGroup{inputParam}{Input parameters for initial data}
\begin{metadataField}{object1}{str}{inputParam}{0}{}
Keyword description to identify the object 1 type.  One of \{\texttt{bh}, \texttt{ns}\}.
\end{metadataField}
\begin{metadataField}{object2}{str}{inputParam}{0}{}
Keyword description to identify the object 2 type.  One of \{\texttt{bh}, \texttt{ns}\}.
\end{metadataField}
\begin{metadataField}{initial\_data\_type}{str}{inputParam}{0}{}
Type of initial data.  One of \texttt{BBH\_CFMS} -- conformally flat, maximal slice; \texttt{BBH\_SKS} -- superposed Kerr-Schild; \texttt{BBH\_SHK} -- superposed harmonic Kerr-Schild \cite{Varma:2018sqd}; \texttt{BBH\_SSphKS} -- superposed spherical Kerr-Schild \cite{Chen:2021rtb}; \texttt{BHNS}; \texttt{NSNS}.
\end{metadataField}
\begin{metadataField}{initial\_separation}{float}{inputParam}{0}{}
Coordinate separation $D_0$ between centers of compact objects, as passed to the initial data solver~\cite{Cook:2004kt, Buonanno:2010yk, Ossokine:2015yla} (code units).
\end{metadataField}
\begin{metadataField}{initial\_orbital\_frequency}{float}{inputParam}{0}{}
Initial orbital frequency $\Omega_0$ passed to the initial-data solver~\cite{Buonanno:2010yk, Ossokine:2015yla} (code units).
\end{metadataField}
\begin{metadataField}{initial\_adot}{float}{inputParam}{0}{}
Radial velocity parameter $\dot{a}_0$ passed to the initial data solver~\cite{Buonanno:2010yk, Pfeiffer:2007yz}.
\end{metadataField}

\metadataGroup{IDmeasure}{Measurements of initial data}
\begin{metadataField}{initial\_ADM\_energy}{float}{IDmeasure}{0}{}
ADM energy of the initial data (code units).
\end{metadataField}
\begin{metadataField}{initial\_ADM\_linear\_momentum}{list[float]}{IDmeasure}{0}{}
ADM linear momentum of the initial data (code units).
\end{metadataField}
\begin{metadataField}{initial\_ADM\_angular\_momentum}{list[float]}{IDmeasure}{0}{}
ADM angular momentum of the initial data (code units).
\end{metadataField}
\begin{metadataField}{initial\_mass1}{float}{IDmeasure}{0}{}
Christodoulou mass of apparent horizon 1 at initial data (code units).
\end{metadataField}
\begin{metadataField}{initial\_mass2}{float}{IDmeasure}{0}{}
Christodoulou mass of apparent horizon 2 at initial data (code units).
\end{metadataField}
\begin{metadataField}{initial\_dimensionless\_spin1}{list[float]}{IDmeasure}{0}{}
Dimensionless spin of object 1 in the initial data.
\end{metadataField}
\begin{metadataField}{initial\_dimensionless\_spin2}{list[float]}{IDmeasure}{0}{}
Dimensionless spin of object 2 in the initial data.
\end{metadataField}
\begin{metadataField}{initial\_position1}{list[float]}{IDmeasure}{0}{}
Initial coordinates of the center of body 1.
\end{metadataField}
\begin{metadataField}{initial\_position2}{list[float]}{IDmeasure}{0}{}
Initial coordinates of the center of body 2.
\end{metadataField}

\metadataGroup{referenceParam}{Reference quantities}
\begin{metadataField}{relaxation\_time}{float}{referenceParam}{0}{}
Time at which we deem junk radiation to have sufficiently decayed (code units).
\end{metadataField}
\begin{metadataField}{reference\_time}{float}{referenceParam}{0}{}
Time at which reference quantities are extracted from the evolution (code units).
\end{metadataField}
\begin{metadataField}{reference\_mass1}{float}{referenceParam}{0}{}
Christodoulou mass of black hole 1 at reference time (code units).
\end{metadataField}
\begin{metadataField}{reference\_mass2}{float}{referenceParam}{0}{}
Christodoulou mass of black hole 2 at reference time (code units).
\end{metadataField}
\begin{metadataField}{reference\_dimensionless\_spin1}{list[float]}{referenceParam}{0}{}
Dimensionless spin of object 1 at reference time.
\end{metadataField}
\begin{metadataField}{reference\_dimensionless\_spin2}{list[float]}{referenceParam}{0}{}
Dimensionless spin of object 2 at reference time.
\end{metadataField}
\begin{metadataField}{reference\_position1}{list[float]}{referenceParam}{0}{}
Coordinates of the center of body 1 at reference time.
\end{metadataField}
\begin{metadataField}{reference\_position2}{list[float]}{referenceParam}{0}{}
Coordinates of the center of body 2 at reference time.
\end{metadataField}
\begin{metadataField}{reference\_orbital\_frequency}{list[float]}{referenceParam}{0}{}
Orbital angular frequency vector at reference time (code units).
\end{metadataField}
\begin{metadataField}{reference\_mean\_anomaly}{float}{referenceParam}{0}{}
Mean anomaly at reference time.
\end{metadataField}
\begin{metadataField}{reference\_eccentricity}{float}{referenceParam}{0}{}
Orbital eccentricity at reference time~\cite{Mroue:2010re}.
\end{metadataField}

\metadataGroup{finalProperties}{Properties of merger/final quantities}
\begin{metadataField}{number\_of\_orbits\_from\_reference\_time}{float}{finalProperties}{2}{}
Number of orbits from reference time until formation of a common apparent horizon.
\end{metadataField}
\begin{metadataField}{number\_of\_orbits\_from\_start}{float}{finalProperties}{2}{}
Number of orbits from start of simulation until formation of a common apparent horizon.
\end{metadataField}
\begin{metadataField}{common\_horizon\_time}{float}{finalProperties}{0}{}
Evolution time at which common horizon is first detected.
\end{metadataField}
\begin{metadataField}{remnant\_mass}{float}{finalProperties}{0}{}
Final mass of the remnant black hole after merger.
\end{metadataField}
\begin{metadataField}{remnant\_dimensionless\_spin}{list[float]}{finalProperties}{0}{}
Dimensionless spin of the remnant black hole after merger.
\end{metadataField}
\begin{metadataField}{remnant\_velocity}{list[float]}{finalProperties}{0}{}
Linear velocity of the remnant black hole after merger.
\end{metadataField}

\metadataGroup{codeInfo}{Code information}
\begin{metadataField}{spec\_revisions}{list[str]}{codeInfo}{0}{}
Array of git revisions of the evolution code.
\end{metadataField}
\begin{metadataField}{spells\_revision}{str}{codeInfo}{0}{}
Git revision of initial data solver.
\end{metadataField}
\begin{metadataField}{date\_link\_earliest}{str}{codeInfo}{2}{}
Earliest link time of code used to perform this simulation.
\end{metadataField}
\begin{metadataField}{internal\_changelog}{dict}{codeInfo}{2}{}
Text describing changes made in different \verb|internal_minor_versions| of this local simulation. Always starts empty for new simulations.
\end{metadataField}
\begin{metadataField}{internal\_minor\_version}{int}{codeInfo}{2}{}
Incremented when anything changes in this
local simulation that is not tracked by the fields \verb|metadata_format_revision|,
\verb|metadata_content_revision|, or \verb|postprocess_revision|. No
relation to DOI revision numbers. Always starts at 0 for new
simulations.
\end{metadataField}
\begin{metadataField}{metadata\_content\_revision}{int}{codeInfo}{2}{}
Incremented when values in the metadata change (which should seldom happen). Updated for all (non-deprecated) simulations at once. No relation to DOI revision numbers. At the time of this catalog release, all non-deprecated simulations carried \verb|metadata_content_revision=1|.
\end{metadataField}
\begin{metadataField}{metadata\_format\_revision}{int}{codeInfo}{2}{}
Incremented when keys in the metadata change (which should seldom happen). Updated for all (non-deprecated) simulations at once. No relation to DOI revision numbers. At the time of this catalog release, all non-deprecated simulations carried \verb|metadata_format_revision=2|.
\end{metadataField}
\begin{metadataField}{postprocess\_revision}{int}{codeInfo}{2}{}
Incremented when anything changes that is not a raw SpEC output, such as re-computation of extrapolation, center-of-mass-correction, or memory-correction using newer algorithms or different parameters. Should not occur often. Updated for all (non-deprecated) simulations at once. No relation to DOI revision numbers. At the time of this catalog release, all non-deprecated simulations carried \verb|postprocess_revision=1|.
\end{metadataField}
\begin{metadataField}{t\_relaxed\_algorithm}{dict}{codeInfo}{2}{}
\verb|t_relaxed_algorithm| is a dict. It contains fields:
\begin{itemize}
  \item \verb|algorithm|: either `HHT' or `RMS'.
  \item \verb|reason|: present only for `RMS' method; text explaining why HHT method failed and we fell back to RMS.
  \item \verb|reference_time_method|: usually absent, but
        `\verb|set_by_hand|' for certain simulations (almost head-on)
        where a reference time was explicitly set by hand and is not
        correlated with \verb|relaxation_time|.
\end{itemize}
\end{metadataField}
\begin{metadataField}{pbj\_info}{dict}{codeInfo}{2}{}
This is a dict that contains
\{ \verb|base_lev| : \texttt{str},
   \verb|transition_time| : \texttt{float},
   \verb|base_lev_bitwise_identical| : \texttt{str} \}
\begin{itemize}
  \item \verb|base_lev| is the Lev that is shared between all PBandJ Levs. It is a string like `Lev3'.
  \item \verb|transition_time| is the time at which PBandJ
        happens. That is, before \verb|transition_time|, all the Levs
        should be identical. If there is no PBandJ, then
        \verb|transition_time| is 0.0 and \verb|base_lev| is the same
        as the Lev that the metadata.json file is in.
  \item \verb|base_lev_bitwise_identical| is either the string `true'
        or the string `false'. If `true', this means that the current
        Lev (the one the metadata.json file is in) and its
        \verb|base_lev| actually are bitwise identical up to (approx)
        \verb|transition_time|. The `false' case occurs when someone
        ran an additional Lev at a later time, but \verb|base_lev| had
        been deleted by the sysadmins, so the user reran
        \verb|base_lev| and then did PBandJ to start the new lev. But
        the rerun of \verb|base_lev| is not always bitwise identical
        to the original \verb|base_lev| because something changed
        (timing-based stuff in SpEC, compiler version, libraries on
        the cluster, etc).
\end{itemize}
\end{metadataField}

\metadataGroup{timeStamps}{Time stamps information}
\begin{metadataField}{date\_postprocessing}{str}{timeStamps}{2}{}
Timestamp of the most recent postprocessing of the raw simulation data to compute extrapolated, COM-corrected, memory-corrected waveforms.
\end{metadataField}
\begin{metadataField}{date\_run\_earliest}{str}{timeStamps}{2}{}
Timestamp of when this simulation was started.
\end{metadataField}
\begin{metadataField}{date\_run\_latest}{str}{timeStamps}{2}{}
Timestamp of when the last segment of this simulation started.
\end{metadataField}

\subsection{Deprecated metadata fields}

For completeness, below are the metadata fields that are deprecated and  no
longer available. The ``deprecated'' version marks the first
\texttt{metadata\_format\_revision} in which a metadata field is no longer
available.

\metadataGroup{id}{Identification}
\begin{metadataField}{point\_of\_contact\_email}{str}{id}{0}{2}
Contact information for questions.
\end{metadataField}
\begin{metadataField}{authors\_emails}{list[str]}{id}{0}{2}
List of authors' emails.
\end{metadataField}

\metadataGroup{refs}{References}
\begin{metadataField}{simulation\_bibtex\_keys}{list[str]}{refs}{0}{2}
References which should be cited if this simulation is used.
\end{metadataField}
\begin{metadataField}{code\_bibtex\_keys}{list[str]}{refs}{0}{2}
List of bibtex keys which are references about the evolution code.
\end{metadataField}
\begin{metadataField}{initial\_data\_bibtex\_keys}{list[str]}{refs}{0}{2}
List of bibtex keys which are references about the initial data code.
\end{metadataField}
\begin{metadataField}{quasicircular\_bibtex\_keys}{list[str]}{refs}{0}{2}
List of bibtex keys which are references about creating quasicircular initial data.
\end{metadataField}

\metadataGroup{finalProperties}{Properties of merger/final quantities}
\begin{metadataField}{number\_of\_orbits}{float}{finalProperties}{0}{2}
Number of orbits until formation of a common apparent horizon.  Replaced by \verb|number_of_orbits_from_reference_time| and \verb|number_of_orbits_from_start|.
\end{metadataField}

\metadataGroup{codeInfo}{Code information}
\begin{metadataField}{metadata\_version}{int}{codeInfo}{0}{2}
This field has been replaced by the fields \verb|metadata_format_revision| and \verb|metadata_content_revision|.  The 2013 catalog~\cite{Mroue:2013xna} implicitly carried metadata version 0. The 2019 catalog~\cite{Boyle:2019kee} carried metadata version 1.
\end{metadataField}

\section{Algorithm for superseding simulations}
\label{sec:superseded_simulations}

Since we are now deprecating simulations that we believe are not
trustworthy, some automated way of finding new, similar simulations
must be available.  The \software{sxs} package provides the following
algorithm as the default for finding a new simulation that is similar
to the requested, deprecated simulation---although a different
algorithm can be passed as an option.  Our approach is very
simplistic: we measure the distance between simulations in terms of
their metadata parameters at their respective reference times.  Since
the SXS catalog also contains some simulations with neutron stars,
systems of different types of compact objects are considered
infinitely far apart in parameter space.  We call the two systems we
are comparing $s$ and $s'$.  We compare their mass ratios and
dimensionless spin vectors, as well as their eccentricities.  Because
of the varying thresholds involved in eccentricity reduction, we set a
pair of thresholds on the eccentricity magnitudes, $\epsilon=10^{-2}$
and $\epsilon'=10^{-3}$, as well as a single length threshold $N'=20$.
We ignore the eccentricities (essentially, setting $e=e'=0$) if
\begin{enumerate}
\item the reference eccentricity of system $s$ is below $\epsilon$,
\item the reference eccentricity of system $s'$ is below $\epsilon'$, and
\item the length of system $s'$ is greater than $N'$ orbits.
\end{enumerate}
We choose two different thresholds because our eccentricity-reduction
methods have improved over the years~\cite{Pfeiffer:2007yz,
Buonanno:2010yk, Mroue:2012kv, Habib:2024soh}, and the first system
will generally be an older simulation in which eccentricities as high
as $10^{-2}$ would have been considered non-eccentric, whereas newer
simulations will need to have eccentricity below $10^{-3}$ to be
considered non-eccentric, and thus a good match for the first system.

Finally, we compute the distance by summing the differences of all
parameters in quadrature and taking the square-root:
\begin{equation}
  \label{eq:supersede metric}
  \delta m = \sqrt{
    \left(q-q'\right)^2
    + \left|\vec{\chi}_1-\vec{\chi}_1'\right|^2
    + \left|\vec{\chi}_2-\vec{\chi}_2'\right|^2
    + \left|e \exp[i\ell] - e' \exp[i\ell']\right|^2
  },
\end{equation}
where $q$ and $q'$ are the reference mass ratios of the two systems,
$\vec{\chi}_1$, $\vec{\chi}_1'$, $\vec{\chi}_2$, and $\vec{\chi}_2'$
are the reference dimensionless spins of the two black holes in the
two systems, $e$ and $e'$ are the reference eccentricities, and $\ell$
and $\ell'$ are the reference mean anomalies. Users may pass
\texttt{auto\_supersede=True} to the \texttt{sxs.load} function to
load the ``closest'' simulation by this measure, or choose a threshold
such as \texttt{auto\_supersede=0.01} to raise an exception if no
system with $\delta m \leq 0.01$ can be found. It is also possible to
define a custom \texttt{MetadataMetric} and pass that as an argument
to \texttt{sxs.load}.

\section{Waveform format}
\label{sec:waveform_format}

Data published by the SXS collaboration prior to this publication
occupies more than 12~TiB of storage on the open-access Zenodo
repository---which currently constitutes over 1\% of the total data
storage on Zenodo~\cite{Zenodo}.  This is a problematic amount of data
for a single project, already inducing resistance from the maintainers
of Zenodo.  This paper more than doubles the number of simulations to
be hosted, making the situation untenable.  Limiting the types of data
we publish will reduce the size of the catalog significantly, but
still not enough.

The SXS collaboration has developed a new waveform format that
compresses the data by an average factor of 7 compared to the old
format, which used standard compression.  This new format applies a
number of complicated non-standard transformations to the data.  The
\software{sxs} package provides an interface that insulates the user
from the details of the waveform format, and even allows for seamless
compatibility between different versions of the catalog that use
different formats.  Nonetheless, details of the new waveform format
and comparison to older formats may be of interest.

\begin{figure}
  \includegraphics{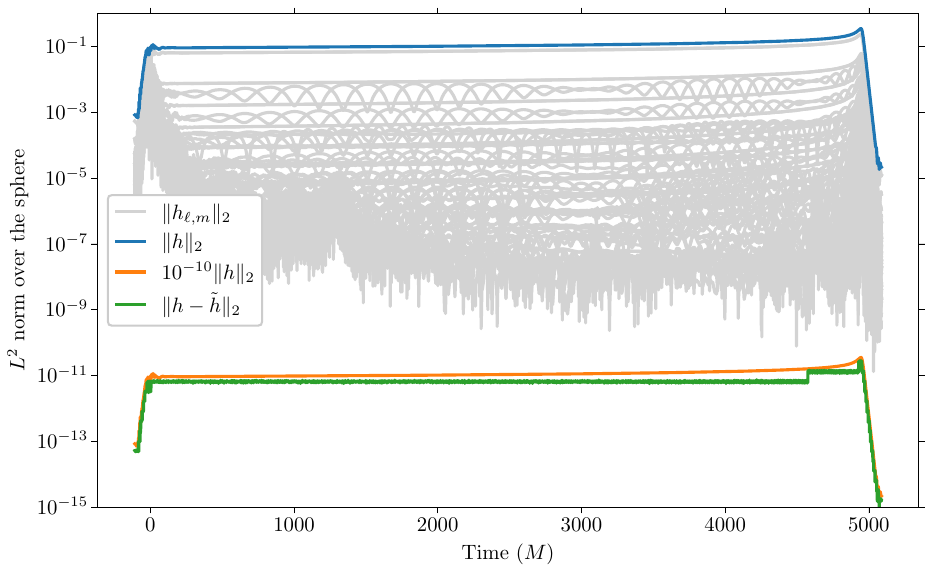}
  \caption{\label{fig:compression_error_paper}%
  Error in the strain data after compression for SXS:BBH:2019---a
  mildly precessing system with mass ratio $q=4$.  The light gray
  curves illustrate the amplitudes of the individual modes of the
  original strain.  Their individual values are unimportant; we simply
  wish to indicate the range of values.  The upper blue curve shows
  the $L^2$ norm of the strain over the sphere at each instant of
  time.  The smooth orange curve towards the bottom shows $10^{-10}$
  times that norm; the $L^2$ norm of the error in the compressed
  waveform is guaranteed to be less than this value.  Finally, the
  green curve at the very bottom shows the actual $L^2$ norm of the
  difference between the original and compressed waveforms.  We see
  that this difference does satisfy the error bound at all times.  It
  changes in steps of roughly a factor of 2 as additional bits in the
  output data can be truncated as the norm increases or must be
  included as the norm decreases, as described in
  Sec.~\ref{sec:RPDMB_truncation}.  The ragged band along the lower
  extent of the gray region suggests that the true error in each mode
  is \emph{at least} that large.  The conservative error tolerance in
  the compressed waveform is chosen to be significantly smaller out of
  concern for the archival integrity of the data.
  }
\end{figure}

The fundamental idea is to manipulate the data in ways that increase
the number of runs of constant bytes, which can  be efficiently
compressed with a standard compression algorithm.  There are several
lossless transformations for which the original data can be
precisely reconstructed, as well as one lossy step with a selectable
tolerance.  The final step is compression with BZIP2~\cite{BZIP2},
followed by storage in a standard HDF5~\cite{HDF5} file.  The full
sequence of transformations is described Sec.~\ref{sec:RPDMB}.  The
results are shown in Fig.~\ref{fig:compression_error_paper},
demonstrating that the $L^2$ norm of the difference between the
original and compressed data is less than $10^{-10}$ times the norm of
the original data at each instant of time.  The resulting files
are---on average---roughly 7x smaller than they would be if the raw
data were stored without compression, and over 5x smaller than
they would be using standard compression techniques of HDF5.

\subsection{Existing waveform formats}
\label{sec:existing_waveform_formats}

Because they represent the precious output of relatively expensive and
long-running numerical-relativity simulations, waveform data have long
been stored in formats designed to ensure that there is no information
loss.  The state of the art for lossless compression, used by most numerical relativity
groups, is to store the data in HDF5 files with standard shuffle and
gzip filters to enable compression~\cite{HDF5}.\footnote{It is
important to note that ``chunking'' must also be done properly to
ensure that the shuffle and gzip filters can perform adequately.  In
short, the data should be arranged so that successive timesteps of a
single mode are contiguous in memory, so that data values change
slowly as memory is traversed.} However, this limits the amount of
compression that can be achieved.  As more and more waveforms appear
in catalogs, storing and distributing the data in lossless formats
becomes increasingly difficult.  Lossy formats have also been used,
where a tradeoff is made between the accuracy of the data and the size
of the file~\cite{Aasi:2014tra, Ajith:2012az, GridRefinement,
Brown2007, Schmidt:2017btt, RomSpline}.  As numerical relativity catalogs become more
unwieldy, this becomes a more attractive option.

The NINJA2 project~\cite{Aasi:2014tra, Ajith:2012az} created one of
the earliest instances of a collection of waveforms from various numerical relativity
groups, including ``hybridized'' data extended with post-Newtonian
waveforms.  One need that was identified was for a common format that
could drastically reduce the size of the data, so that waveforms could
be distributed and analyzed more easily.  Specifically, because of the
varied requirements of the many groups involved, the format needed to
be based on plain text files, eliminating the possibility of applying
standard compression algorithms.  Instead, the \software{GridRefinement}
code~\cite{GridRefinement} was introduced to reduce the number of time
steps stored, while still allowing the original data to be
reconstructed to within a specified tolerance.  The essential idea was
to decompose the waveform modes into amplitude and ``unwrapped'' phase
(with branch-cut discontinuities removed through the addition of
multiples of $2\pi$), then reduce the number of time steps to store
such that the original data could be reconstructed to within a
specified tolerance via \emph{linear} interpolation.  Linear
interpolation was chosen only for simplicity; restrictions of the
project required that the code be self-contained and simple enough
that it would not require review.  By default, the tolerances were
$10^{-5}$ for the (instantaneous) relative amplitude error and the
absolute phase error.  The algorithm broke up each time series
into one or more time intervals.  First, the entire time series
was chosen as a single interval.
If the data could not be reconstructed to
within the specified tolerance, the midpoint between the first and
last time steps was included, and the process repeated on each of the
two new intervals, applied recursively until the data could be
adequately reconstructed.  The amplitude-phase decomposition was
chosen because that early work only considered non-precessing systems,
which---because of their symmetry---exhibit smooth variations in amplitude
and phase.  This would no longer be as useful for the precessing
systems in more modern catalogs.  Nonetheless, for the waveforms
produced for the NINJA-2 project, this approach reduced the size of
waveform files by ``anywhere from a few percent for short numerical
data to 99\% for very long hybrid waveforms.''~\cite{Brown2007}

The LVCNR format~\cite{Schmidt:2017btt} extends the NINJA-2 format
by using spline
interpolation instead of linear.  While it retains the amplitude-phase
decomposition, it includes an option to store real and imaginary parts
if the result would be smaller.  It also uses a different tolerance,
error measure, and algorithm~\cite{RomSpline} to determine the time
steps to store.  This format is still used by much of the
LIGO-Virgo-KAGRA Collaboration for exchanging waveforms.  For
comparison, when loosening the tolerance of our new RPDMB format
(described below in Sec.~\ref{sec:RPDMB})
to achieve the same
accuracy as the default LVCNR values (which is achieved using $\tau =
10^{-5}$), files created by LVCNR---even after eliminating all but the
essential waveform data---are roughly 15 times larger than those
created by RPDMB.\footnote{This is true for LVCNR files without the
metadata---such as spins and orbital elements as functions of
time---and with the \texttt{slim=True} option passed to eliminate the
extraneous ``error'' data, so that \emph{only} the minimal set of
waveform data are included in the LVCNR file.  Also note that one
limitation of the reference implementation of the LVCNR format is that
it scales poorly---typically as the cube of the number of time steps
in the waveform.  Because the waveforms presented here are relatively
long compared to those the LVCNR format was designed for, converting
just the public waveforms presented in this catalog would take an
estimated 20,000~CPU-hours.  Using an algorithm closer to the one used
by the \software{GridRefinement} code, we can achieve equivalent results
about 4,000 times faster.  This improvement has been implemented in
\texttt{sxs.utilities.lvcnr}, but is not used for these comparisons
to ensure fidelity to the reference implementation, though the results
would be essentially identical.}

A notable feature of these lossy formats is that some data points are
simply dropped from the data, while those that remain are stored in
unaltered form.  This stands in contrast to the RPDMB format, which
stores all data points, but modifies them for more effective
compression.

\subsection{Compressing in RPDMB format}
\label{sec:RPDMB}

Here, we outline in more detail the steps involved in converting data
to the new RPDMB format.  We considered and took inspiration from a number
of sources dealing with compression of data~\cite{EngelsonEtAl2000,
LIGOCompression2001, LindstromIsenburg2006, Ratanaworabhan2006,
MAFISC2013, MASUI2015181, SPDP2018}.  Note that the complete process
is implemented as the \texttt{sxs.rpdmb.save} function, while
\texttt{sxs.rpdmb.load} will load the data back into the original
form.

RPDMB stands for \texttt{rotating\_paired\_diff\_multishuffle\_bzip2},
which describes each element of the format \textit{per se}.  However,
we include two additional steps in the process of converting waveforms
to this format: truncation and adding zero---the first of which is
probably the single most important operation for actually reducing the
size of the file.  The following are all the steps of the conversion
process, in order.

\subsubsection{Corotating frame}
\label{sec:RPDMB_corotating_frame}

The older waveform formats mentioned in
Sec.~\ref{sec:existing_waveform_formats} decomposed the waveform into
amplitude and phase to take advantage of the fact that these
quantities vary on a secular timescale for non-precessing systems,
unlike the real and imaginary parts which vary on the orbital
timescale.  For more general systems, the phase in particular can vary
almost discontinuously in time, making this a poor choice.  However,
even for precessing systems we can factor out the sinusoidal
dependence of the real and imaginary parts of the waveform modes by
transforming to a corotating frame.  This is defined as a rotating
frame in which the time dependence of the waveform modes is
minimized~\cite{Boyle:2013nka}.  The angular velocity of this frame
can be easily computed from the modes themselves, and that velocity
can then be integrated to provide the frame as a function of
time~\cite{Boyle_2017}.  The waveform is then transformed into that
frame.  The entire procedure is implemented as the
\texttt{to\_corotating\_frame} method to be applied to
\texttt{sxs.WaveformModes} objects.

This has the effect of making the real and imaginary parts of the
waveform modes vary on a slower timescale; for non-precessing systems,
they vary at the same rate as the amplitude.  However, we have also
introduced another piece of data that we need to store for each
waveform: the frame itself.  The calculations use a quaternion
representation, in which the frame is represented by four numbers at
each instant of time.  However, these numbers are not independent; the
sum of their squares must be 1.  Instead, we can use a more compact
representation: the logarithm of the quaternion.  This is a
three-vector at each instant of time, representing the
\emph{generator} of the rotation (roughly its axis-angle form), rather
than the rotation itself.  The original rotation can be reconstructed
exactly simply by exponentiating the generator, as implemented in the
\software{quaternionic} package~\cite{quaternionic}, and the waveform
rotated back to the inertial frame, as implemented in the
\software{spherical} package~\cite{spherical}.

\subsubsection{Paired modes}
\label{sec:RPDMB_paired_modes}

In the corotating frame, the modes of a non-precessing system will
vary on a secular timescale, but the modes of a precessing system will
still vary on the orbital timescale.  This is because of spin-orbit
coupling causing an asymmetry in the emission of GWs
across the orbital plane.  We can further factor out the asymmetric
waves into symmetric and antisymmetric parts, each of which again
varies on a secular timescale~\cite{Boyle:2014ioa}.  This is done
simply by combining mode $(\ell, m)$ with the conjugate of mode
$(\ell, -m)$, with an appropriate sign.  For simplicity, we define the
sum and difference:
\begin{equation}
  s^{\ell, m} = \frac{h^{\ell, m} + \bar{h}^{\ell, -m}}{\sqrt{2}},
  \qquad
  d^{\ell, m} = \frac{h^{\ell, m} - \bar{h}^{\ell, -m}}{\sqrt{2}}.
\end{equation}
Which of these is symmetric and which is antisymmetric depends on the
value of $\ell$, but is irrelevant for our purposes; we only care that
both are slowly varying.  We then define the collective quantity
\begin{equation}
  f^{\ell,m} = \begin{cases}
    d^{\ell, -m} & m < 0, \\
    h^{\ell, 0} & m = 0, \\
    s^{\ell, m} & m > 0.
  \end{cases}
\end{equation}
Note that the transformation from $h$ to $f$ is reversible, up to
machine precision, via
\begin{equation}
  h^{\ell,m} = \begin{cases}
    \frac{\bar{f}^{\ell, -m} - \bar{f}^{\ell, m}}{\sqrt{2}} & m < 0, \\
    f^{\ell, 0} & m = 0, \\
    \frac{f^{\ell, m} + f^{\ell, -m}}{\sqrt{2}} & m > 0.
  \end{cases}
\end{equation}
It is implemented as the \texttt{convert\_to\_conjugate\_pairs} method
in the \software{sxs} package, and the resulting $f^{\ell,m}$ is passed
on to the next step.

\subsubsection{Truncation}
\label{sec:RPDMB_truncation}
  
As is standard in computation, \software{SpEC} relies primarily on
64-bit floating-point numbers to represent the data---specifically,
the IEEE 754 \texttt{binary64} format---which is accurate to a
relative precision of roughly $10^{-16}$.  However, due to the nature
of numerical evolutions, the results of \software{SpEC} simulations are generally
accurate to significantly fewer bits.  Thus, many of the
lowest-significance bits in the waveform are effectively random.  This
means that they are essentially incompressible, and yet contribute no
useful information.  By discarding this randomness in some way, we can
improve the overall compression of the waveforms without any real loss
to the information content of the data.

The most obvious approach would be to simply use a numerical
representation with lower precision---for example by using
\texttt{binary32} instead of \texttt{binary64}.  However, this would
be a very coarse approach, with no flexibility to adapt to the data,
and uniform \emph{relative} precision (around $10^{-7}$ for
\texttt{binary32}) for all modes at all times.  For example, in an
equal-mass non-precessing system, the largest modes will generally be
$(\ell, m) = (2, \pm 2)$, while the $(8, \pm 1)$ modes are typically 9
orders of magnitude smaller.  There would seem to be little reason to
store the $(8, \pm 1)$ modes to a precision of $10^{-7}$, when even
its most significant digits are smaller than the least significant
digits of the $(2, \pm 2)$ modes.  Moreover, the \emph{relative}
magnitude of various modes will generally change significantly over
the course of a simulation.  These points suggest two important
criteria for truncation:
\begin{enumerate}
  \item different modes should have different (relative) precision, and
  \item the precision should depend on time.
\end{enumerate}
Neither of these are satisfied by a simple change of floating-point
format.

The HDF5 specification~\cite{HDF5Docs} includes ``N-Bit'' and
``scale-offset'' filters to provide more precise control over the
precision of the data.  The N-Bit filter allows the user to specify
the number of bits of the data to be stored, which in principle would
allow us to simply ignore bits below a certain significance threshold.
The scale-offset filter allows the user to specify an \emph{absolute}
precision, below which the data will be rounded to zero.  However,
both of these features must be specified on a per-dataset basis,
leaving no possibility to adapt to time dependence in the data.

We can easily implement a generalization of these filters that allows
us to specify a time-dependent precision.  Using the fact that our
floating-point numbers are specified in binary form, we can multiply
by an appropriate power of 2, then round the result to the nearest
integer, and then divide by the same power of 2.  This is equivalent
to setting bits in the binary representation of the number to zero
below a certain significance.\footnote{We assume that the data are
sufficiently well behaved that we can ignore any subtleties with
subnormal numbers or infinities.}  We can compute the $L^2$ norm $n_i$
of the field at each time $t_i$.  Then, with a tolerance $\delta$
\emph{relative to that norm}, we find the value of the least-significant bit
greater than or equal to $\delta n_i$, and construct $p_i$, the
smallest power of 2 such that $p_i$ times that bit's value will be
greater than or equal to 1.
We then multiply each mode by $p_i$ and round the result to the nearest integer,
then divide again by $p_i$ to get the truncated number:
\begin{gather}
  n_i = \sqrt{\int_{S^2} \left|h(t_i, \theta, \phi) \right|^2
  \, d\Omega} = \sqrt{\sum_{\ell,m} \left|h^{\ell,m}(t_i)\right|^2}
  = \sqrt{\sum_{\ell,m} \left|f^{\ell,m}(t_i)\right|^2},
  \\
  p_i = 2^{\left\lfloor{-\log_2\left(\delta\,n_i\right)}\right\rfloor},
  \\
  \mathring{f}^{\ell,m}(t_i) = \frac{\text{round}\left(f^{\ell,m}(t_i) \,
  p_i\right)}
  {p_i}.
\end{gather}
Because we have used an exact power of 2, the division in the last
line will be exact in \texttt{binary64}, with the result that the
binary representation of $\mathring{f}^{\ell,m}(t_i)$ will have all bits
below the significance threshold set to zero.  Those zeros can be
compressed very effectively, especially after application of the
``multishuffle'' step described below.  This is implemented as the
\texttt{truncate} method in the \software{sxs} package.

Here, $\delta$ describes the worst-case error in each component (real
or imaginary) of each mode of the $f^{\ell,m}$ data at each instant of
time.  Denoting the number of modes as $N_{\textrm{modes}}$, the
worst-case error in the \emph{total waveform} at each instant will
obey
\begin{equation}
  \sqrt{\sum_{\ell,m} \left|
    h^{\ell,m}(t_i) - \mathring{h}^{\ell,m}(t_i)
  \right|^2}
  \leq \delta \sqrt{N_{\textrm{modes}}}.
\end{equation}
Thus, if we have in mind some total error tolerance $\tau$, we need to
set the per-component tolerance $\delta$ as
\begin{equation}
  \delta = \frac{\tau} {\sqrt{N_{\textrm{modes}}}}.
\end{equation}
For the waveforms published as part of this catalog, we chose $\tau =
10^{-10}$, limiting the individual error at each instant of time in
each component of each of the 77 modes to roughly $1.14 \times
10^{-11}$ times the norm of the waveform at that instant.  This choice
was made simply by plotting the amplitude of all modes and judging the
level at which numerical errors become visually obvious, exhibiting
discontinuities and noise.  We expect that this choice is several
orders of magnitude more conservative than necessary, because of the
numerous and cumulative errors inherent in evolving
numerical-relativity data.  Nonetheless, to ensure the archival
quality of the data, we chose to err on the side of caution.

\begin{figure}
  \includegraphics{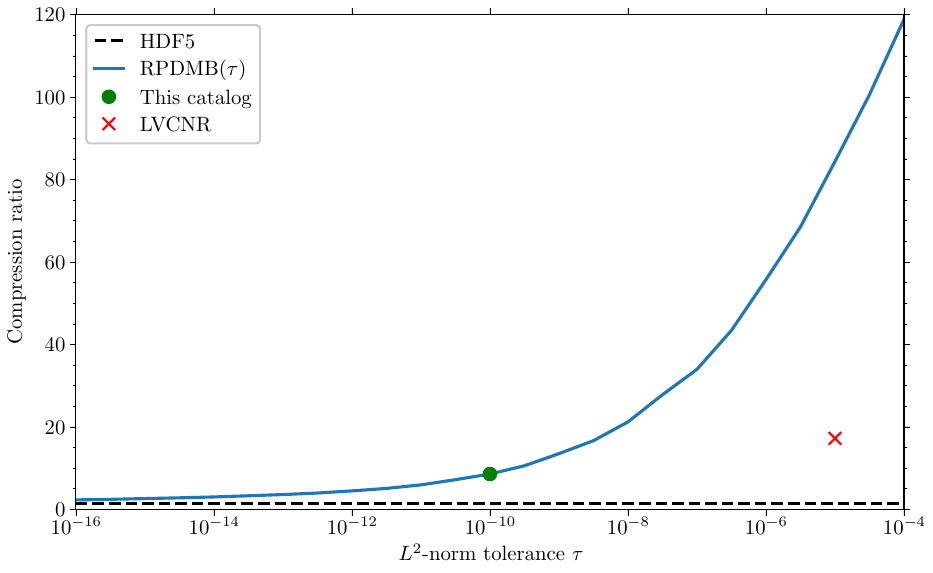}
  \caption{ \label{fig:compression_ratio_v_tolerance} %
  Compression ratio relative to the size of the raw data (waveform
  modes and time) as a function of tolerance $\tau$ for the strain
  from SXS:BBH:2265, which was the last simulation in the previous SXS
  catalog.  The horizontal dashed line shows the compression ratio
  1.3x when using standard HDF5 storage, storing one chunk per real
  time series, applying the shuffle and gzip filters.  The blue curve
  shows the compression ratio achieved by RPDMB using various levels
  of truncation $\tau$, with the green dot emphasizing the tolerance
  $\tau=10^{-10}$ used for the data described by this paper---the
  compression ratio for this particular waveform being 8.6x.  The red
  ``x'' shows the compression ratio 17x achieved by the LVCNR format,
  excluding metadata and other non-waveform information.  Note that
  the LVCNR tolerance is specified as $10^{-6}$, but uses a different
  error measure; we compare it here to $\tau=10^{-5}$, which is
  sufficient for RPDMB to reproduce the waveform to higher accuracy by
  either error measure with a compression ratio of 84x.
  }
\end{figure}

The particular choice of $\tau$ has a significant effect on the
compression ratio.  For example, we can examine the effect of changing
$\tau$ on the compression ratio for a particular waveform.
Figure~\ref{fig:compression_ratio_v_tolerance} shows the compression
ratio as a function of $\tau$ for the strain from SXS:BBH:2265, which
was the last simulation in the previous SXS catalog.
A compression ratio of 1 would be equivalent to storing the exact
bytes of the data.  The horizontal dashed line shows the compression
ratio, 1.3x, when storing in HDF5 using the best available standard
options, but without otherwise altering the data in any way.  The
solid blue curve shows the compression ratio achieved by RPDMB using
various levels of truncation.  Note that even with $\tau=10^{-16}$,
\emph{smaller} than machine precision, RPDMB can still achieve better
compression than HDF5 because there are modes with significant bits
smaller than machine precision relative to the norm of the waveform;
these are truncated, resulting in a reduction in compressed file size.

As the tolerance is increased, the compression ratio increases.  For
this particular waveform, a compression ratio of 8.6x is achieved with
the default value of $\tau=10^{-10}$ used in the public data.
However, with less stringent tolerances, the compression ratio grows
drastically.  In particular, this growth is much faster than one would
expect simply from counting the number of nonzero bits to be
stored---ranging from a factor of 2 improvement at small $\tau$ to an
order of magnitude at large $\tau$.  Presumably, this is because the
bits being zeroed out are effectively random and therefore cannot be
compressed, whereas the remaining bits are increasingly continuous and
compress relatively well.

The red ``x'' shows the compression factor, 17x, achieved by the LVCNR
format with its default settings, including only the time and waveform
modes, and excluding the various spin and orbital time series and
other metadata.  By default, the LVCNR format specifies a tolerance of
$10^{-6}$, where the error is given as the maximum of the absolute
value of the difference between the original data and the data
reconstructed by a degree-5 interpolating spline~\cite{RomSpline},
applied separately to the amplitude and phase of each mode or the real
and imaginary parts of each mode---whichever one results in a smaller
set of points.  This differs from the error measure used in this
paper, as we compute the error at each instant of time rather than
taking the maximum, and measure it relative to the norm of the
waveform.  If the norm at time $t_i$ is $n_i$, then we can achieve a
comparable accuracy to the LVCNR format with a tolerance of $\tau
\approx 10^{-6} / n_i$.  By choosing $\tau=10^{-6} / n_\mathrm{max}$,
where $n_\mathrm{max}$ is the maximum norm of the waveform, we can
ensure that the error in the RPDMB waveform is always less than the
error in the LVCNR waveform by \emph{both} error measures.  For the
waveform shown in Fig.~\ref{fig:compression_ratio_v_tolerance}, this
corresponds to $\tau = 10^{-5}$.  The actual compression ratio
achieved by LVCNR with its default tolerance is shown at this value of
$\tau$ as a red ``x'' mark, with a value of 17x.  For comparison,
RPDMB achieves a compression ratio of 84x at this tolerance, while
achieving smaller errors by either measure.

\subsubsection{Adding zero}
  
At this stage, many of the paired modes will be precisely zero.
Specifically, for non-precessing simulations, the amplitude of half of
the mode pairs should zero by symmetry; they will generally not be
exactly zero due to numerical error, but after truncation they likely
will be.  However, the standard floating-point representation of zero
is not unique; floating-point zeros can have either sign.  Because
those zeros were generated by---effectively---random noise, their
signs will also be effectively random.  This adds a great deal of
entropy to the data, making it difficult to compress.
  
We can improve the compression by setting all zero-valued data points
to a unique representation of zero.  Fortunately, the IEEE
floating-point standard~\cite{IEEEFloatingPoint2019} specifies that
adding two zeros of either sign will result in a positive zero (by
default).  Thus, we can simply add zero to all the data to eliminate
the sign ambiguity.  Including this step reduces the total size of the
waveform data by an average of $0.5\%$.  Though this is a tiny
improvement, it is achieved by a trivial operation, which can be
combined with the truncation step above at essentially no cost.  It
should be noted that this is almost surely only beneficial because of
the truncation step and the presence of very small modes.  As more
modes are included, there will be more very small modes, making them
more likely to be truncated to zero, and thus making this step more
beneficial---though likely always quite small.

\subsubsection{Differencing sequential data points}
  
For reasonably continuous data, as in a time series, we can improve
compression by storing the first data point as is, but thereafter only
the differences between sequential data points.  Specifically, we
store $h^{\ell,m}(t_0)$ as is, and then store $h^{\ell,m}(t_j) -
h^{\ell,m}(t_{j-1})$ for all $j > 0$.  Note that subtraction here is
based on floating-point numbers; we could also reinterpret the numbers
as 64-bit integers and perform integer subtraction, which would lead
to different patterns in the resulting bits.  A closely related
procedure reinterprets the numbers as \emph{unsigned} 64-bit integers,
then uses \XOR in place of subtraction.  Note that the \XOR operation
is precisely invertible, whereas floating-point subtraction can accrue
roundoff-level errors.  For data that have already been subject to
truncation those are negligible, whereas \XOR is required for data
that have not been truncated.\footnote{These are both well known
techniques, used in many compression
algorithms~\cite{EngelsonEtAl2000, LIGOCompression2001,
LindstromIsenburg2006, Ratanaworabhan2006, MAFISC2013, MASUI2015181,
SPDP2018}.  Sometimes referred to as ``delta encoding'', both can be
considered simple cases of the more general technique of ``predictive
encoding'', which uses other data points to predict a given value, and
then stores only the difference between the prediction and the true
value.}

We have tested the effectiveness of both forms of differences and of
\XOR when applied to the waveform data.  The results are similar
but---at least in combination with all other steps described
here---floating-point differencing achieves 37\% better compression
than integer differencing and 60\% better compression than \XOR when
storing the real and imaginary parts of the waveform modes and the
components of the logarithm of the rotation quaternion.  However, to
retain lossless compression for the time data, we use \XOR for that
part of the data.

\subsubsection{Multishuffle}  The HDF5 library includes a ``shuffle''
filter~\cite{HDF5} that reorders the bytes of the data to improve
compression for continuous data.  Conceptually, we can imagine storing
a series of numbers
\begin{displaymath}
  12345, 12346, 12347, 12348.
\end{displaymath}
If each numeral is stored in its obvious order---the order in which it
appears above---the pattern-finding techniques underlying many
compression algorithms will not be able to take advantage of the fact
that the numbers are very similar.  However, if the numbers are stored
as
\begin{displaymath}
  11112222333344445678,
\end{displaymath}
those longer runs of repeated digits can be encoded more efficiently.
This is the basic idea behind the shuffle filter.  When applied to
much longer sequences of 64-bit numbers, the effect can be quite
dramatic.  Specifically, the ``shuffle'' filter takes the
most-significant byte (8 bits) of each number and stores them
consecutively, followed by the second-most-significant byte, and so
on.  Another way to think of this is to consider a series of $N$
64-bit numbers as a matrix of $N$ rows and 8 columns, where each
column represents a byte of a number.  The obvious storage scheme
would be to go across each row, then down to the next row, and so on.
The shuffle filter is essentially a transpose---it goes down all rows
in the first column, then the second column, and so on.

It might be helpful to describe this more specifically as a
``byteshuffle'' filter, because this approach can be generalized to
put it on a spectrum of similar filters.
Reference~\cite{MASUI2015181} introduced the ``bitshuffle'' filter,
which takes the most-significant bit of each number and stores them
consecutively, followed by the second-most-significant bit, and so on.
Again, we could think of this as a matrix transpose, except that the
columns now are individual bits.  We can easily imagine using
different divisions also.  The ``nibbleshuffle'' would use groups of 4
bits (nibbles), and the ``morselshuffle'' would use groups of 2 bits.
(We will see below that groups of sizes that are not powers of 2 are
strongly disfavored.)

Each of these options has its advantages and disadvantages, depending
on the scale across which the bits can be expected to vary coherently.
We would expect byteshuffle to be the best choice when consecutive
bytes correlate well with each other; bitshuffle would be best when
that correlation fails at the byte level but holds at the bit level.
However, because of the nature of the \texttt{binary64} format, where
different bits have different significance, we would expect the
optimal choice to vary along the length of the 64 bits.  This suggests
that the best choice could be to vary the width of the groups of bits
that we shuffle as we progress along the number.  We call this
``multishuffle''; it is implemented via the function of the same name
in the \software{sxs} package.

A multishuffle filter must be specified by the widths of the groups of
bits that are shuffled.  For example, if we group the first byte of
each number, followed by the next 4 bits, followed by the next 2 bits,
and then individual bits after that, we would specify the filter as
$(8, 4, 2, 1, 1, \ldots)$.  The sum of all those numbers must be 64.
In this representation, no shuffling would be $(64)$, the standard
byteshuffle is $(8, \ldots)$, nibbleshuffle is $(4, \ldots)$,
morselshuffle is $(2, \ldots)$, and bitshuffle is $(1, \ldots)$.

In general, there are $2^{63} \approx 10^{19}$ possible multishuffle
filters for 64-bit data, so the search is not trivial.  Because of the
discrete nature of the problem, and the enormous size of the space of
possibilities, a genetic algorithm seems like a natural choice for
finding the best filter.  We used the \software{Evolutionary.jl}
package~\cite{Evolutionary_jl} to search for the shuffle widths that
delivered the best compression for a random subset of 20 simulations,
using each waveform from the various resolutions and extrapolation
orders, for a total of 472 waveforms.  Any form of shuffling was
always better than not shuffling at all, by at least 30\%.  However,
among all other choices tested, the bitshuffle algorithm was the
worst, followed by morselshuffle, then the standard byteshuffle, and
nibbleshuffle.  Nonetheless, it was possible to gain significant
improvements by varying the shuffle widths---by 18\% to 3\% over the
various fixed-width options, and in particular about 6.5\% over the
standard byteshuffle.

At no point throughout the optimization did the genetic algorithm find
better results for a multishuffle involving a width that was not a
power of 2.  This is presumably because the compression algorithm
(BZIP2, as described below) is still based on bytes, so combining
groups of bits with sizes that are not powers of 2 will not produce
alignment along those bytes that can be easily compressed.  The
algorithm very quickly determined that the first groups should have
sizes 16, 4, 2, 2, and 2.  Beyond that, the results clearly favored
small groups of either 1 or 2 bits, but did not depend very strongly
on the exact choice of widths.  The widths we have used for the
waveform data in this catalog are
\begin{equation}
  \label{eq:shuffle_widths}
  (16, 4, 2, 2, 2, 1, 1, 1, 1, 1, 1, 1, 1, 1, 1, 1, 1, 1, 1, 1, 1, 1, 1, 1, 1, 1, 1, 4, 4, 4, 4)
\end{equation}
For \texttt{binary64}, the first 16 bits represent the sign, the
exponent, and the first 4 bits of the significand.  The next 4 bits
represent the next 4 bits of the significand, and so on, with all
remaining bits representing decreasingly significant bits.

Recalling that the previous step converted the data to successive
differences, we are really encoding the \emph{rate of change} of the
data between timesteps.  In particular, the first block of 16 bits
will essentially encode the difference between successive timesteps,
rounded to just the first few digits.  The rate of change in the
modes---at least to this level of accuracy---will typically be quite
small, so we might expect that many of these values will be repeated,
meaning that run-length encoding will efficiently compress the data.
This turns out to be true for the first 8 bits, but not for all 16
bits.  Instead, it appears that the benefit of grouping the first 16
bits together is that there are relatively few distinct patterns that
appear in the data.  Of the $2^{16}=65,536$ possible values that could
be encoded by the first 16 bits, only 1,000 or so actually appear in
typical waveforms, with the most common values occurring \emph{far}
more often than most.  This is ideal for the
Huffman-coding~\cite{Huffman1952} stage of compression to represent
these values more compactly than the 16 bits would.\footnote{Extending
beyond 16 bits, this particular advantage disappears; the number of
observed values grows with the number of possible values.  In that
case, it is evidently better to take advantage of other features of
the compression.}  These features are decreasingly likely to occur as
we proceed to bits with lower significance, which is why it makes
sense that the shuffle widths decrease as we proceed through the
number.  On the other hand, there is an increasing likelihood that the
bits will switch to being a series of 0s, which can be compressed very
effectively with run-length encoding.  We want those runs to occur as
soon as possible, which is why it makes sense that widths become 1 as
that condition becomes more likely.  Finally, the last 4 groups of 4
bits represent the last 16 bits of the significand, which will always
be less significant than the $10^{-10}$ tolerance.  Therefore, we can
expect that these are just long runs of 0s for every mode.  The
genetic algorithm found essentially no difference in any choice of
widths for these bits, so we group them together to reduce the number
of times the data must be traversed.

It is important to note that the optimal choice of shuffle widths may
vary for different datasets; the results we report here are true for
our data, after the particular processing steps described above.  They
may not extend to different types of datasets, or even to the same
waveforms when preceded by different processing steps.  As such, the
precise set of shuffle widths is essentially adjustable, and therefore
should be stored as metadata alongside the compressed data.  As
mentioned below, we choose the HDF5 file format to organize the data,
and preserve the shuffle widths as an HDF5 attribute of each dataset.
When loading the data, this attribute is read to ensure that the data
are decompressed correctly.

One caveat to note about all types of shuffle filters is that they
require multiple passes over the data.  The number of passes required
is set by the length of the width specification: 8 for byteshuffle, 64
for bitshuffle, and 31 for the widths chosen for this catalog.  In the
naive application, this can lead to memory bottlenecks with numerous
cache misses.  When this is a problem, the data can be divided into
``chunks'' that fit into cache, and the shuffle filter applied to each
chunk separately.  In particular, the reference implementation of the
bitshuffle filter~\cite{MASUI2015181} uses this technique.  This will
reduce the number of cache misses, but will also reduce the
effectiveness of the compression stage.  Because we find the cost to
be minimal compared to the increased burden of disk access due to
less-effective compression, we choose to treat each real or imaginary
part of each waveform mode as a single chunk to be shuffled as a unit.

\subsubsection{BZIP2}

None of the preceding stages actually reduce the number of bytes that
must be stored.  In fact, the conversion to a rotating frame just adds
another set of data---the logarithm of the quaternions---to be stored.
Instead, each stage has been designed to reduce the entropy of the
information in the waveform.  The final stage is to use that reduced
entropy to compress the data.  We arrange the data as a single
sequence of bytes, beginning with the shuffled time data, followed by
the shuffled modes, and finally the shuffled logarithm of the rotation
quaternions.

We have tested a variety of standard compression algorithms to do so.
The best results were obtained using BZIP2~\cite{BZIP2}, which passes
the data through a number of stages, including run-length encoding,
the Burrows-Wheeler transform~\cite{BurrowsWheeler1994}, the
move-to-front transform~\cite{Ryabko1980, Bentley1986}, and Huffman
coding~\cite{Huffman1952}.  Files created using XZ/LZMA~\cite{xz} were
about 1\% larger; about 3\% for Brotli~\cite{brotli}; 12\% for
GZIP~\cite{gzip}; and 13\% for Zstd.

The speed of compression varied widely---from an order of magnitude
faster for Zstd, to an order of magnitude slower for Brotli.  However,
the time spent compressing the data is generally a small fraction of
the time spent transforming and writing the data, so the speed of
compression is not a significant concern.  The speed of decompression
was less varied, but also dominated entirely by the time spent reading
the data from disk---not to mention the time spent transferring the
data over the internet.  Thus, the dominant factor in choosing a
compression algorithm was the size of the compressed file, leaving
BZIP2 as the clearly preferred choice.

\subsubsection{HDF5 storage}

The output of the BZIP2 compression stage is a single byte stream,
which could be written to disk directly.  However, there are also
various pieces of metadata that are important for being able to
reliably decompress the data.  For example, the shuffle widths used
for the multishuffle filter must be stored, as must the number of
modes and/or the number of time steps.  Perhaps most importantly, to
ensure that the files remain useful in the future and that the
interface can be easily extended, it is important to store the name of
the format used to compress the data.  It can also be helpful to
organize multiple waveforms into a single file.

The HDF5 file format is well suited to this task, and is widely used
in the scientific community for storing large datasets along with
metadata~\cite{HDF5}.  Specifically, we store the data stream as a
single ``opaque'' dataset named ``data'', with no filters of any kind.
We then attach ``attributes'' to that dataset:
\begin{itemize}
  \item \texttt{sxs\_format}
  \item \texttt{shuffle\_widths}
  \item \texttt{ell\_min}
  \item \texttt{ell\_max}
  \item \texttt{n\_times}
\end{itemize}
The first attribute value is
\texttt{rotating\_paired\_diff\_multishuffle\_bzip2}, which allows the
interface to automatically detect the correct decompression steps to
apply.  For all waveforms in this catalog, the shuffle widths are as
given in Eq.~\eqref{eq:shuffle_widths}, and the minimum and maximum
values of $\ell$ are 2 and 8, respectively.  The number of time steps
is also stored.  While this could be inferred from the length of the
data after BZIP2 decompression, it is convenient to store it directly,
acting as a simple check on the integrity of the data.  This dataset
and its attributes can be stored in any group of an HDF5 file, whether
the root group as for \texttt{Strain\_N2.h5} files, or in descriptively
named subgroups as in the \texttt{ExtraWaveforms.h5} files.

\subsection{Future work}

By including the name of the format within the file itself, we have
left open the possibility of changing the format in the future.  There
may be better choices for the multishuffle widths, or for the
compression algorithm used in the final stage.  There is almost surely
some improvement that could be made in the predictive step.  For
example, the various modes could be normalized in some way, or make
use of post-Newtonian approximations.  However, it is not clear that
such a change would be worth the effort.  Surely the most impactful
alteration would be to simply increase the tolerance $\tau$ used in
the truncation step.  Regardless, all of these considerations will
remain almost invisible to the user, because the \software{sxs}
interface will be
able to detect the format and apply the correct decompression steps
automatically.

\section{Simulations with large differences in
  Figure~\ref{fig:Extrapolation_order_and_Psi4_constraint_comparison}}
\label{sec:simul-with-large}

In Figure~\ref{fig:Extrapolation_order_and_Psi4_constraint_comparison}
there are long tails at large waveform difference that are caused by a
handful of simulations, which we list here for completeness.

For the plots comparing extrapolation orders, the problematic simulations
are head-on, nearly-head-on, or scattering cases.
These simulations are particularly short and hence a large fraction 
of the waveform still contains a considerable amount of initial transient junk 
radiation. The differences between extrapolation orders are always relatively
large during these initial transients, but due to the shortness 
of these particular runs their relative contribution to the waveform difference 
is larger. Some of these runs also have waveform-extraction radii that
are too close together, which causes unusually large extrapolation errors,
especially for high extrapolation order.  These simulations are:
\begin{multicols}{4}
  \begin{itemize}
\item[] SXS:BBH:1110
\item[] SXS:BBH:1363
\item[] SXS:BBH:1544
\item[] SXS:BBH:3873
\item[] SXS:BBH:3874
\item[] SXS:BBH:3875
\item[] SXS:BBH:3876
\item[] SXS:BBH:3877
\item[] SXS:BBH:3879
\item[] SXS:BBH:3881
\item[] SXS:BBH:3882
\item[] SXS:BBH:3883
\item[] SXS:BBH:3884
\item[] SXS:BBH:3885
\item[] SXS:BBH:3887
\item[] SXS:BBH:3889
\item[] SXS:BBH:3890
\item[] SXS:BBH:3995
\item[] SXS:BBH:3996
\item[] SXS:BBH:3997
\item[] SXS:BBH:3999
\item[] SXS:BBH:4000
\item[] SXS:BBH:4292
  \end{itemize}
\end{multicols}

\section*{Affiliation list}
\markboth{{\scshape Affiliation list}}{}
\addcontentsline{toc}{section}{Affiliation list}
\begingroup
\let\clearpage\relax
\affilInfo{} %
\endgroup

\section*{References}
\markboth{{\scshape References}}{}
\addcontentsline{toc}{section}{References}
\bibliographystyle{iopart_num}
\bibliography{catalog-202X}

\end{document}